\documentclass[default]{sn-jnl}

\jyear{2022}%

\theoremstyle{thmstyleone}%
\theoremstyle{thmstyletwo}%

\theoremstyle{thmstylethree}%

\usepackage{amsmath}
\usepackage{amsfonts}
\usepackage{bm}
\usepackage{makecell}
\usepackage{multirow}
\usepackage{graphicx}
\usepackage{subfigure}
\usepackage{float}
\usepackage{comment}

\newcommand{\xic}[1]{{\color{black} #1}}

\newcommand{\ve}[1]{\mbox{\boldmath ${#1}$}}
\newcommand{\vesub}[2]{\mbox{{\boldmath ${#1}$}$_{#2}$}}

\begin{document}

\title[Chen et al.]{Designing Compact Features for Remote Stroke Rehabilitation Monitoring using Wearable Accelerometers}

\author[1]{\fnm{Xi} \sur{Chen}}\email{chenxi@hainanbank.com.cn}

\author[2]{\fnm{Yu} \sur{Guan}}\email{yu.guan@warwick.ac.uk}

\author[3]{\fnm{Jian Qing} \sur{Shi}}\email{shijq@sustech.edu.cn}

\author[4]{\fnm{Xiu-Li} \sur{Du}}\email{duxiuli@njnu.edu.cn}

\author[5]{\fnm{Janet} \sur{Eyre}}\email{janet.eyre@ncl.ac.uk}

\affil[1]{
\orgname{Hainan Rural Credit Union}, 
\state{Hainan}, \country{China}}

\affil[2]{\orgdiv{Department of Computer Science}, \orgname{University of Warwick}, 
\country{UK}}

\affil[3]{\orgdiv{Department of Statistics $\&$ Data Science,}, \orgname{Southern University of Science $\&$ Technology}, 
\country{China}}

\affil[4]{\orgdiv{School of Mathematical Sciences}, \orgname{Nanjing Normal University}, 
\country{China}}

\affil[5]{\orgdiv{Institute of Neuroscience}, \orgname{Newcastle University}, 
\country{UK}}


\abstract{Stroke is known as a major global health problem, and for stroke survivors it is key to monitor the recovery levels. However, traditional stroke rehabilitation assessment methods (such as the popular clinical assessment) can be subjective and expensive, and it is also less convenient for patients to visit clinics in a high frequency.
To address this issue, in this work based on wearable sensing and machine learning techniques, we \xic{develop} an automated system that can predict the assessment score in an objective manner. With wrist-worn sensors, accelerometer data \xic{is} collected from 59 stroke survivors in free-living environments for a duration of 8 weeks, and we \xic{map} the week-wise accelerometer data (3 days per week) to the assessment score by developing signal processing and predictive model pipeline.
To achieve this, we \xic{propose} two types of new features, which can encode the rehabilitation information from both paralysed and non-paralysed sides while suppressing the high-level noises such as irrelevant daily activities. Based on the proposed features, we further \xic{develop} the longitudinal mixed-effects model with Gaussian process prior (LMGP), which can model the random effects caused by different subjects and time slots (during the 8 weeks). Comprehensive experiments \xic{are} conducted to evaluate our system on both acute and chronic patients, and \xic{the promising results suggest} its effectiveness.}


\maketitle



\maketitle

\section{Introduction}\label{section:intro}
It is widely known that stroke is a worldwide health problem causing disability and death \cite{Donnan:2008}, and it occurs when a blood clot cuts off oxygen supply to a region of the brain.
Hemiparesis is a very common symptom of post-stroke that is the fractional or intact paralysis of one side of the body, i.e., the opposite side to where the blood clot \xic{occurs}, and it results in difficulties in performing activities, e.g., \xic{with} reduced arm movement.
Patients can recover some of their capabilities with intense therapeutic input, so it is important to assess their recovery levels in time.
There are many approaches to assess patients' recovery levels including brain imaging \cite{BrainImaging_2005}, questionnaire-based \cite{Questionaires_2007}, and lab-based clinical assessment \cite{CAHAI_2005}.

 The brain imaging technique, is deemed as one of the most reliable approach, which can provide the information of brain hemodynamics \cite{BrainImaging_2005}. However, this approach requires special equipment and is very expensive in cost.
 Questionnaire-based approaches investigate the functional ability during a period using questionnaires, and it can be categorised into two types: patient-completed and caregiver-completed \cite{Questionaires_2007}.
 Although it is much cheaper than brain imaging approaches, it may contain high-level of bias. For instance, patients may not remember their daily activities (i.e.,recall bias); the caregivers may not be able to observe the patient all the time. These biases make questionnaire-based approaches less precise.
 Lab-based clinical assessment approaches \cite{CAHAI_2005}\cite{Barreca:2005}, on the other hand, provide an alternative solution.
 The patients' upper limb functionality will be assessed by clinicians, e.g., by observing patients' capabilities of finishing certain pre-defined activities \cite{CAHAI_2005}.
 Compared with braining imaging or questionnaire-based approaches, the cost of lab-based clinical assessment approaches is reasonable with high accuracy.
 However, this assessment is normally taken in clinics/hospitals, which is not convenient for the patients, making continuous monitoring less feasible.

 In this work, we aim to build an automated stroke rehabilitation assessment system using wearable sensing and machine learning techniques.
 Different from the aforementioned approaches, our system can measure the patients objectively and continuously in free-living environments.
 We \xic{collect} accelerometer data using wrist-worn accelerometer sensors, and \xic{design} compact features that can capture rehabilitation-related movements, before mapping these features to clinical assessment scores (i.e., the model training process).  The trained model can be used to infer recovery-level for other unknown patients. 
 In free living environments, there are different types of movements which may be related to different frequencies.
 For example, activities such as running or jumping may correspond to high-frequency signals, while sedentary or eating may be low-frequency signals.
 In this study, instead of recognising the daily activities explicitly, which is hard to achieve given limited annotation (e.g., without frame/sample-wise annotation), we \xic{transform} the raw accelerometer data to the frequency domain, where we design features that can encode the rehabilitation-related movements.
 Specifically, wavelet transform \cite{waveletbook} \xic{is} used, and the wavelet coefficients can represent the particular frequency information at certain decomposition scales.
 In \cite{Preece:2009}, Preece et al. \xic{provide} some commonly used wavelet features extracted from accelerometer data.
 However, to capture stroke rehabilitation-related activities, some domain knowledge should be taken into account to design better features.
 After stroke, patients have difficulties in moving one side (i.e., paralysed side) due to the brain injury, and data from paralysed side tends to describe more about the upper limb functional ability, than the non-paralysed side (i.e., normal side).
 However, such signals can be significantly affected by personal behaviours or irrelevant daily activities, and such noises should be suppressed before developing the predictive models.
 Various wavelet features were studied, and we \xic{propose} two new types of daily-activity-invariant features that can encode information from both paralysed/non-paralysed sides, before developing predictive models for stroke rehabilitation assessment.
 Specifically, in this work our contributions can be summarised as follows:

 \begin{itemize}
 \item \textbf{Stroke-rehab-driven Features}: We \xic{propose} two new types of compact wavelet-based features that can encode information from both paralysed and non-paralysed sides to represent upper limb functional abilities for stroke rehabilitation assessment. It can significantly suppress the influences of personal behaviours or irrelevant daily activities for data collected in the noisy free-living environment.
 \item \textbf{Automated Assessment System}: Based on the proposed stroke-rehab-driven features, we developed the automated system by using the longitudinal mixed-effects model with Gaussian process prior (LMGP). Various predictive models were studied, and we \xic{find} LMGP can model the random effects caused by the heterogeneity nature among subjects in a 8-week longitudinal study.
 \item \textbf{Comprehensive Evaluation}: Comprehensive experiments \xic{are} designed to study the effectiveness of our system. We comprehensively studied the feature subset on modelling the mixed-effects of LMGP. Compared with other approaches, the results \xic{suggest} the effectiveness of the proposed system on both acute and chronic patients.
 \end{itemize}
. 
\section{Background and Related Work}\label{section:background}

As described in Sec.\ref{section:intro}, 
lab-based clinical assessment \xic{is} one of the most effective stroke rehabilitation assessment methods.
In this section, we introduce the lab-based approach named Chedoke Arm and Hand Activity Inventory (CAHAI) scoring \cite{Barreca:2006b}, based on which our automated system can be developed.
Some sensing and machine learning techniques for automated health assessment are also \xic{reviewed in this section}.

\begin{figure}[H]
	\centering
\includegraphics[width=7cm,height=5.5cm]{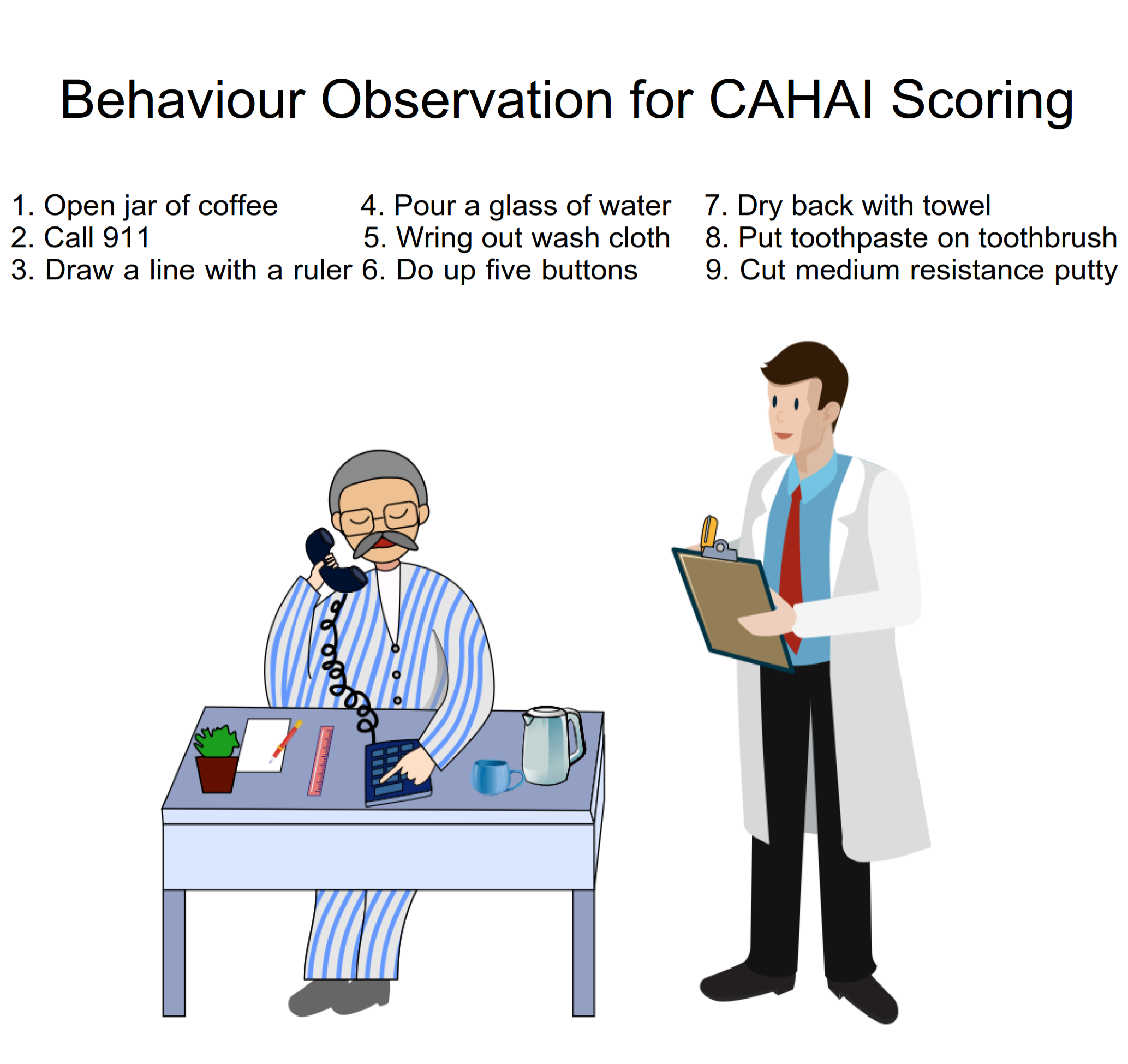}
	\caption{The clinical behaviour assessment for CAHAI scoring \cite{Barreca:2006b}. }
    \label{figure:CAHAI1}
\end{figure}
\subsection{Chedoke Arm and Hand Activity Inventory (CAHAI)}
CAHAI scoring is a clinical assessment method for stroke rehabilitation, and
it is a fully validated measure \cite{Barreca:2006b} of upper limb functional ability with 9 tasks which are scored by using a 7-point quantitative scale.
In the assessment, the patient will be asked to perform 9 tasks, including opening a jar of coffee, drawing a line with a ruler, calling 911, etc. and the clinician will score these behaviours based on patient's performance at a scale from 1 (total assist weak) to 7 (complete independence i.e., timely, safely) \cite{Barreca:2006b}. A task example "call 911" is shown in Fig. \ref{figure:CAHAI1}.
Thus the minimum and maximum summation scores are 7 and 63 respectively.
A CAHAI score form can be found in Fig.\ref{figure:CAHAI2}  in Appendix \ref{section:CAHAI_form}.

\subsection{Automated Behaviour Assessment using Wearables}
Recently, wearable sensing and machine learning (ML) techniques \xic{are} comprehensively studied for automated health assessment.
Compared with the traditional assessment approaches (e.g., via self-reporting, clinical assessment, etc.) which are normally subjective and expensive, the automated systems may provide an objective, low-cost alternative, which can also be used for continuous monitoring/assessment.
Some automated systems \xic{are} developed to assess the behaviours of diseases such as
Parkinson's disease \cite{PD_Rehman19} \cite{Nils_PD15},
autism \cite{TP_autism}, depression \cite{Little_depression2020}; or to monitor the health status such as sleep \cite{Bing_sleep} \cite{Yike_sleep}, fatigue \cite{Yang_ISWC}, \cite{Fatigue_IMU} or recover-level from surgery \cite{Anna_IMU} \cite{gait_surgery}, etc.

After collecting behaviour or physiological signals (e.g., accelerometers, ECG, audio, etc.), assessment/monitoring models can be developed.
For application with high interpretability requirement, feature engineering can be a crucial step.
For example, with gait parameters extracted from IMU sensors (such as stride, velocity, etc.), one can build simple ML models (e.g., random forest) for Parkinson's disease classification \cite{PD_Rehman19} or fatigue score regression \cite{Fatigue_IMU}.
Compared with the redundant IMU data, gait parameters are more compact and interpretable, making it suitable for clinical applications.
However, designing interpretable/clinically-relevant features can be a time-consuming process, which may also require domain knowledge \cite{Bing_sleep}\cite{Fatigue_IMU} \cite{PD_Rehman19}\cite{Anna_IMU}  \cite{gait_surgery}.


On the other hand, when interpretability is less required, deep learning can be an alternative approach, which can be directly applied to the raw signal \cite{Yike_sleep} or engineered features \cite{Nils_PD15} \cite{Bing_sleep} \cite{Yang_ISWC} \cite{Little_depression2020} for (high-level) representation learning and classification/regression tasks.
However, it normally requires adequate data annotation for better model generalisation.

\subsection{Sensing Techniques for Automated Stroke Rehabilitation Monitoring}
With the rapid development of the sensing/ML techniques, researchers also \xic{start} to apply various sensors for stroke rehabilitation monitoring.
In \cite{Dolatabadi:2017}, Kinect sensor \xic{is} used in a home-like environments to detect the key joints such that stroke patients' behaviour can be assessed.
In \cite{sensing_sEMG}, a wireless surface Electromyography (sEMG) device \xic{is} used to monitor the muscle recruitment of the post-stroke patients to see the effect of orthotic intervention.
In clinical environments, five wearable sensors \xic{are} placed on the trunk, upper and forearm of the two upper limbs to measure the reaching behaviours of the stroke survivors \cite{sensing_5sensors}.
To monitor motor functions of stroke patients during rehabilitation sessions at clinics, an ecosystem including a jack and a cube for hand grasping monitoring, as well as a smart watch for arm dynamic monitoring was designed \cite{sensing_ecosystem}.
These techniques can objectively assess/measure the behaviours of the stroke patients, yet they are either limited to clinical environments \cite{sensing_ecosystem}\cite{sensing_5sensors} \cite{sensing_sEMG} or constrained environments (e.g., in front of a camera \cite{Dolatabadi:2017}).

Most recently, wrist-worn sensors \xic{are} used for stroke rehabilitation monitoring for patients in free-living environment \cite{Shane_ISWC} \cite{Tang_stroke}.
In each trial, 3-day accelerometer data \xic{are} collected from both wrists (with a trial-wise annotation, i.e., CAHAI score), and for both works \cite{Shane_ISWC} \cite{Tang_stroke} data analysis \xic{is} performed using the sliding window approach.
To reduce the data redundancy of the raw data, PCA features \xic{are} extracted from each window \cite{Shane_ISWC} \cite{Tang_stroke}.
Moreover, due to the lack of window-wise annotation, in \cite{Shane_ISWC} pseudo label \xic{is} assigned to each window such that a random forest regressor can be trained, while in \cite{Tang_stroke} Gaussian Mixture Models (GMM) clustering approach \xic{is} employed to learn the holistic trial-wise representation, before developing the regression model.
Both methods \cite{Shane_ISWC} \cite{Tang_stroke} \xic{suffer} from the lack of annotation.
In \cite{Shane_ISWC},pseudo labeling \xic{is} introduced, yet the trained model \xic{is} affected by the introduced label noise.
In \cite{Tang_stroke}, the application of GMM clustering (on the sliding windows) \xic{makes} it computationally expensive to large data, and the trained model \xic{does} not generalise well to unseen subjects.

In our work, by analysing the nature of the paralysed/non-paralysed sides, we design stroke-rehab-driven features which can directly encode the long accelerometer sequence (e.g., a trial with 3-day accelerometer data) into a very compact representation.
The features are expected to emphasis the stroke-related behaviours while suppressing the irrelevant activities. 
Based on the proposed features, a predictive model that is adaptive to different subjects/time-slots can be developed using LMGP \cite{Shi:2012} for CAHAI score prediction.

\section{Methodology}\label{section:Methodology}

In this section, we \xic{introduce} our method from data collection, data pre-processing, feature design to predictive models.
Our aim is to develop an automated model which can
map the free-living 3-day accelerometer data into the CAHAI score.
With the trained model, we can automatically infer the CAHAI score in an objective and continuous manner.
To achieve this, we first \xic{reduce} the data redundancy via preprocessing and design compact and discriminant features.
Given the proposed features, a longitudinal mixed-effects model with Gaussian Process prior (LMGP) \xic{is} used \cite{Shi:2012}, which can further reduce the impact of large variability (caused by different subjects and time slots) for higher prediction results.






\subsection{Data Acquisition}\label{section:datacollect}
\begin{figure}[H]
	\centering	\includegraphics[width=10cm,height=5.5cm]{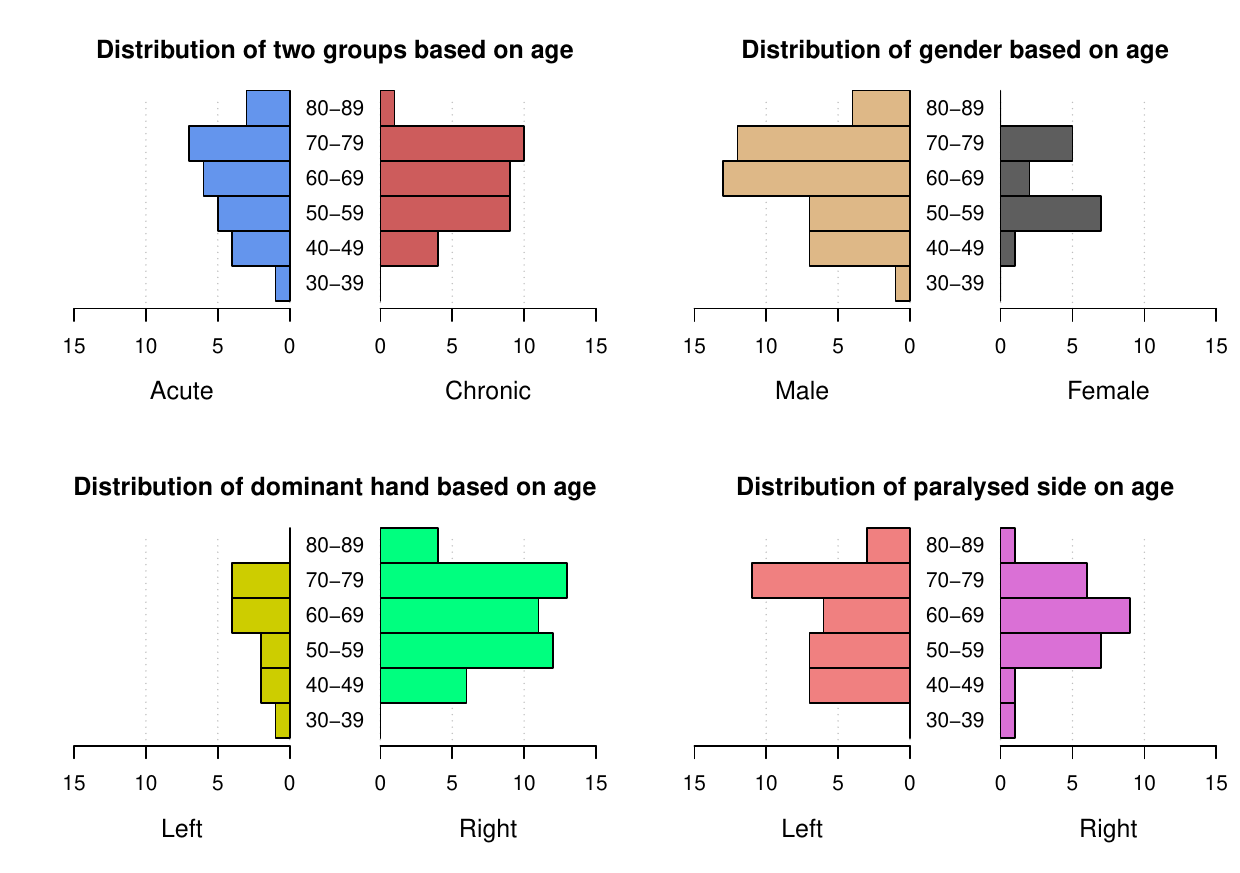}
	\caption{Demographic information of the collected dataset (with 59 subjects): the distributions of acute/chronic condition, gender, dominant/non-dominant hand, paralysed/non-paralysed side with respect to age. }
	\label{figure:distribution}
\end{figure}
\paragraph{Participants}
Data \xic{is} collected as part of a bigger research study which aims to use a bespoke, professionally-written video game as a therapeutic tool for stroke rehabilitation \cite{Shi:2013}. Ethical approval \xic{is} obtained from the National Research Ethics Committee and all work undertaken \xic{is} in accordance with the Declaration of Helsinki. Written, informed consent from all the subjects \xic{is} obtained. A cohort of 59 stroke survivors, without significant cognitive or visual impairment, \xic{are} recruited for the study.
Patients were divided into two groups, i.e.,
\begin{itemize}
\item \textbf{Group 1}: the acute patient group, consisting of 26 participants who enrolled into the study within 6 months after stroke;
\item \textbf{Group 2}: the chronic patient group, \xic{including} 33 participants who were 6 months or more post onset of stroke.
\end{itemize}
The distributions of acute/chronic condition, gender, dominant/non-dominant hand, paralysed/non-paralysed side with respect to age are 
shown in Fig. \ref{figure:distribution}.

These 59 patients \xic{visit} the clinic for the CAHAI scoring every week (a random day in weekdays) for a duration of 8 weeks.
In the 8 weeks, they \xic{are} asked to wear two wrist-worn sensors for 3 full days (including night time) a week.
They \xic{are} also advised to remove the device during shower or swimming.
Since some patients \xic{need} time to get familiar to this data collection procedure, for better data quality we \xic{do} not use the first week's accelerometer data.
The first week's CAHAI scores \xic{are} used as medical history information.

\paragraph{Data collection}
In contrast to other afore-mentioned sensing techniques \cite{sensing_5sensors}\cite{sensing_ecosystem}\cite{sensing_sEMG}\cite{Dolatabadi:2017}, in this study we \xic{collect} the accelerometer data from wrist-worn sensors in free-living environments.
The sensor used for this study, i.e., AX3 \cite{Axivity}, is a triaxial accelerometer logger that \xic{is} designed for physical activity/behaviour monitoring, and it has been widely used in the medical community (e.g., for the UK Biobank physical activity study \cite{ax3_UKB}).
The wrist bands \xic{are} also designed such that the users can comfortably wear it without affecting their behaviours.
The data \xic{is} collected at 100Hz sampling rate, which can well preserve the daily activities of human being \cite{Bouten:1997}.
Different from human activity recognition which requires sample-wise or frame-wise annotation \cite{Guan_HAR17} \cite{HAR_TP}, the data collection in this study is relatively straight-forward.
The patients put on both wrist-worn sensors 3 full days a week, before visiting clinicians for CAHAI scoring (i.e., week-wise annotation).
In other words, we aim to use accelerometer data captured in free-living environments to represent the stroke survivors' upper limb activities to measure the degree of paresis \cite{Henrik:1999} (i.e., CAHAI score).


One problem with most commercial sensors is that only summary data (e.g., step count from fitbit), instead of raw data, are available.
The algorithms of producing summary data are normally non-open source, and may vary from vendor to vendor -- making the data collection and analysis device-dependent, and thus less practical in terms of generalisation and scalability.
The AX3 device used in this study, on the other hand, outputs the raw acceleration information in x, y, z directions.
It is simple and transparent, making the collected data re-usable, which is crucial for research communities.

\subsection{Data pre-processing}\label{section:preprocessSVM}
\begin{figure}[H]
	\centering
	\includegraphics[width=5.5cm,height=4.5cm]{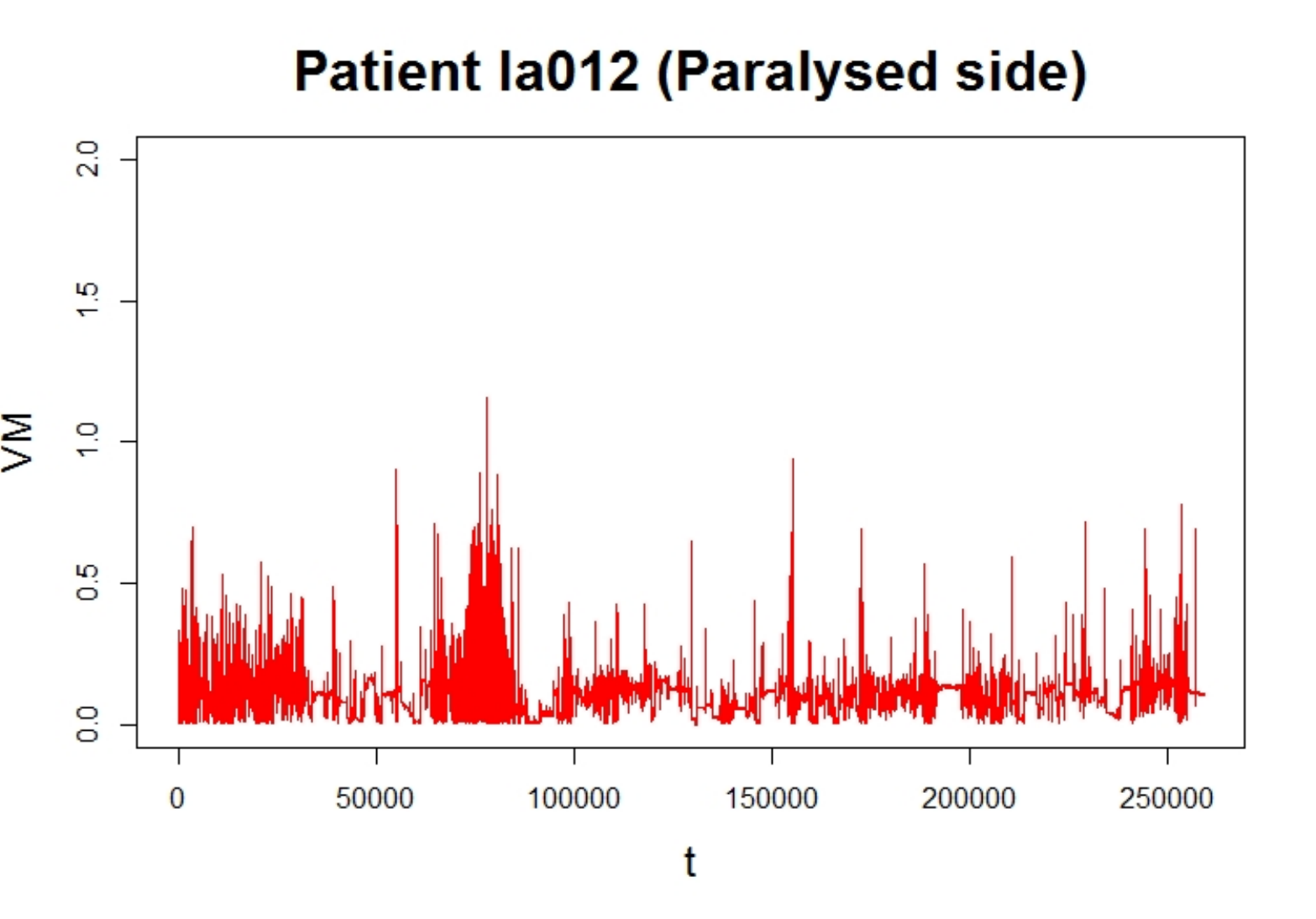}
	\includegraphics[width=5.5cm,height=4.5cm]{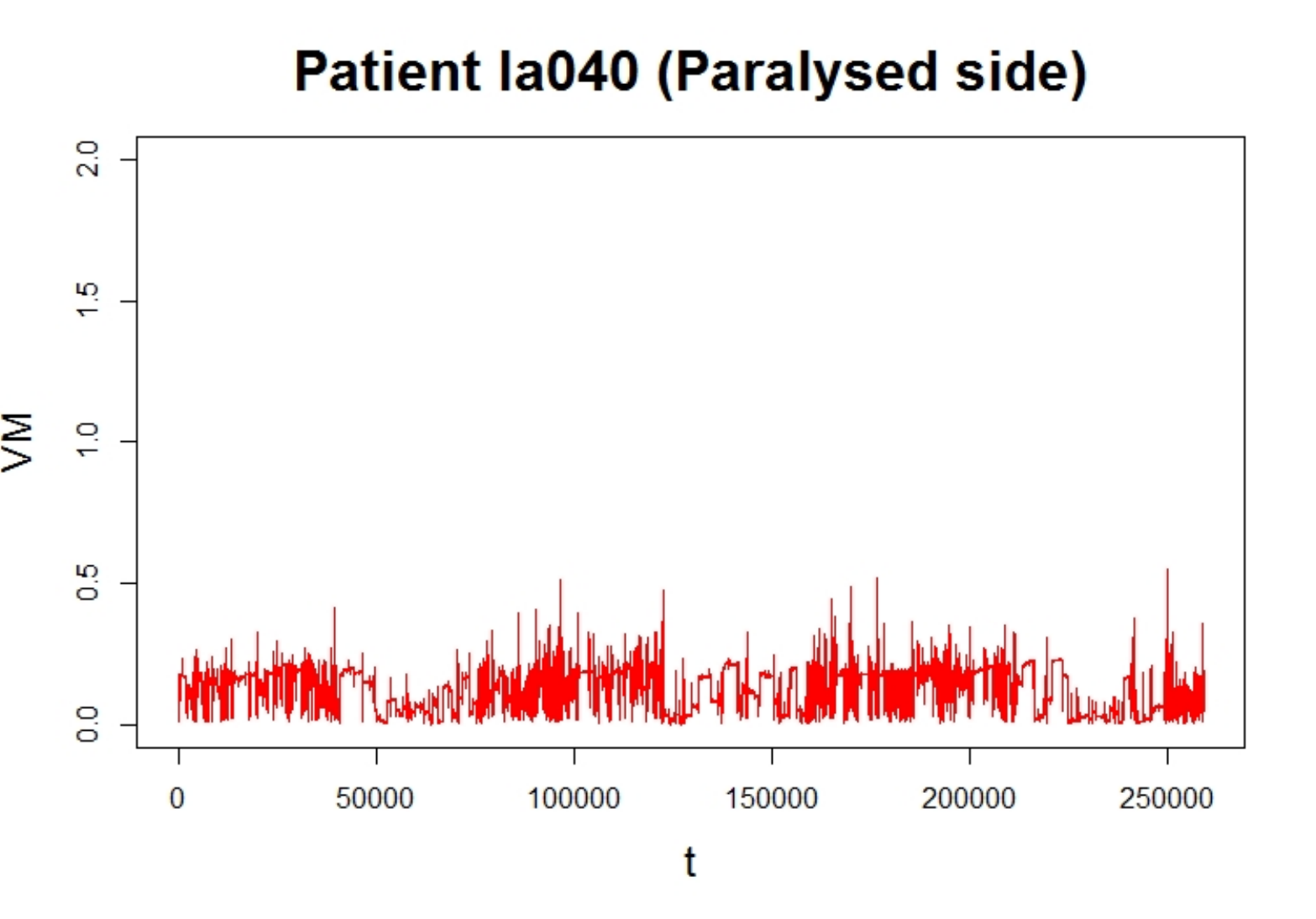}
	\caption{The signal vector magnitude (VM) data collected from two patients (on the paralysed side); Patient la012 has a CAHAI score of 55, while Patient la040 has a CAHAI score of 26.
}
	\label{figure:twoTYPESpatient}
\end{figure}
For accelerometer data, signal vector magnitude (VM) \cite{acc_SVM} is a popular representation, which is simply the magnitude of the triaxial acceleration data defined as
$a(t) =  \sqrt{a_x^2(t) + a_y^2(t) + a_z^2(t)},$
where $a_x(t)$,$a_y(t)$,$a_z(t)$ are the acceleration along the x, y, z axes at timestamp $t$.
The gravity effect can be removed by
$VM(t) =  \lvert a(t) -1  \rvert$.
Because its simplicity and effectiveness, VM has been widely used in health monitoring tasks, such as fall detection \cite{acc_SVM}, physical activity monitoring \cite{ax3_UKB}, perinatal stroke assessment \cite{babystroke}, etc.
To further reduce the data volume, we used second-wise VM, i.e., the mean VM over each second (including 100 samples 
per second) will be used as new representation.
Some second-wise VM examples (from two patients) can be found in Fig. \ref{figure:twoTYPESpatient}.



\subsection{The Proposed Stroke-Rehab-Driven Features}\label{section:feat_design}
\subsubsection{Challenges}
We aim to build a model that can map the 3-day time-series data to the CAHAI score.
Different from other wearable-based behaviour analysis tasks (e.g.,\cite{TP_autism}\cite{Guan_HAR17}), the annotation here is inadequate.
Even if we used the second-wise VM data, each trial still included roughly $3$ days $\times$ $24$h/day $\times$ $3600$s/h $=259200$ samples (a.k.a. timestamps) with one annotation (i.e., CAHAI score).
In contrast to the popular deep learning based human activity recognition approaches, which can be trained when with rich annotations (in frame-wise or sample-wise level), the lack of annotation makes it hard to learn effective representation directly (using machine/deep learning) from the raw data.
Moreover, since the data \xic{is} collected in free-living environments, and the 3 full days (per week) can be taken in weekdays or weekends, which may increase the intra-subject variability significantly, making it hard to model.
To address the afore-mentioned issues, domain knowledge driven
feature engineering may play a major role in extracting compact and discriminant signatures.

\subsubsection{Wavelet Features}
For time-series analysis, wavelet analysis is a powerful tool to represent various aspects of non-stationary signals such as trends, discontinuities, and repeated patterns \cite{wavelet_IMU}  \cite{waveletbook} \cite{Preece:2009}, which is especially useful in signal compression or noise reduction.
Given its properties, wavelet features have been widely used in accelerometer-based daily living activity analytics \cite{wavelet_IMU}.
In this work, we \xic{use} discrete wavelet transform (\textbf{DWT}) and discrete wavelet packet transform (\textbf{DWPT}) as feature extractors, based on which new features were designed to preserve the stroke rehabilitation-related information.
More details of \textbf{DWT} and \textbf{DWPT} can be found at Appendix \ref{section:DWTDWPT}.

After applying the \textbf{DWT} and \textbf{DWPT}, VM signals can be transformed to the wavelet coefficients at different decomposition scales.
\xic{In this work, \textbf{DWT} coefficients at scales $\{2, 3, 4, 5, 6, 7\}$ and \textbf{DWPT} at scales $\{1.1, 1.2, 1.3, 1.4\}$ are employed, and the corresponding normalised Sum of Absolute value of the coefficients at different Decomposition scales  
(referred to as \textbf{SAD} features) are used as new representation. 
Specifically, \textbf{SAD} includes \textbf{DWPT features} defined as 
	\begin{equation} \label{eq:sad_dwpt}
	\begin{split}
	  \ & SAD_{1.1} = \frac{\left \| \textbf{W}_{3.4} \right \|_1}{N/{2^3}} = 2^3 \frac{\left \| \textbf{W}_{3.4} \right \|_1}{N}, \\
	 \ & SAD_{1.2} = \frac{\left \| \textbf{W}_{3.5} \right \|_1}{N/{2^3}} = 2^3 \frac{\left \| \textbf{W}_{3.5} \right \|_1}{N}, \\
	 \ & SAD_{1.3} = \frac{\left \| \textbf{W}_{3.6} \right \|_1}{N/{2^3}} = 2^3 \frac{\left \| \textbf{W}_{3.6} \right \|_1}{N}, \\
	 \ & SAD_{1.4} = \frac{\left \| \textbf{W}_{3.7} \right \|_1}{N/{2^3}} = 2^3 \frac{\left \| \textbf{W}_{3.7} \right \|_1}{N}, \\
	\end{split}
	\end{equation}
and \textbf{DWT features} defined as
 	\begin{equation} \label{eq:sad_dwt}
  SAD_{j} = \frac{\left \| \textbf{W}_{j} \right \|_1}{N/{2^j}} = 2^j \frac{\left \| \textbf{W}_{j} \right \|_1}{N}, \qquad \qquad j = 2,3,4,5,6,7, \\
	\end{equation}
where $\textbf{W}$ presents the wavelet coefficients 
and $N$ presents the length of the VM data. 
More technical details of $\mathbf{DWT}$, $\mathbf{DWPT}$, as well as the scale selection can be found in Appendix \ref{section:waveletfeatures}.}

\begin{figure}[H]
	\centering
	\includegraphics[width=6.5cm,height=4.5cm]{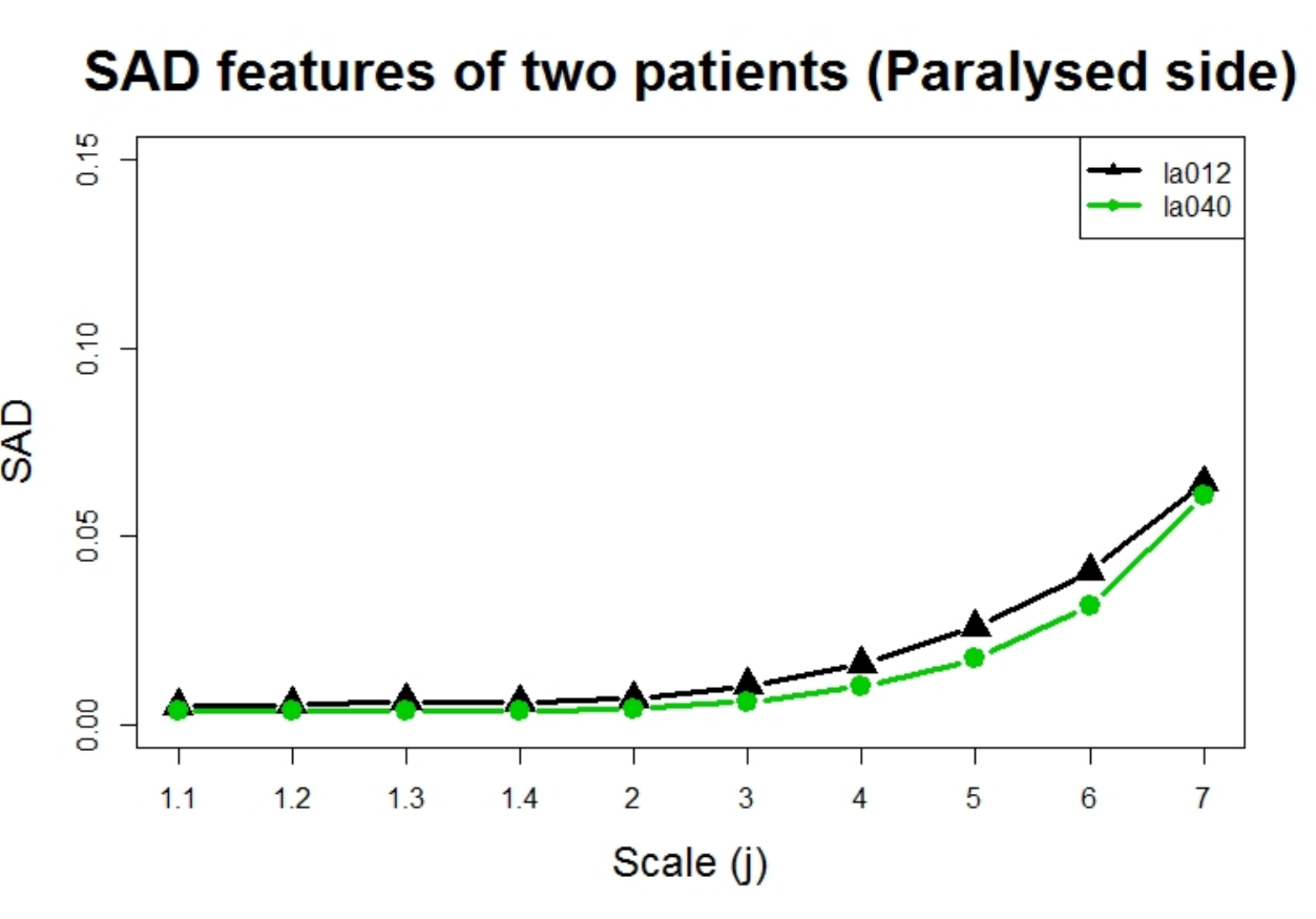}
	\caption{$10$-dimensional $\mathbf{SAD}$ features extracted from the paralysed side of two patients (with different CAHAI scores); They exhibit similar patterns, indicating the necessity of developing more informative stroke-related features.}
	\label{figure:SADla012la038P}
\end{figure}

Through wavelet transformation, the long sequence (e.g., VM data in Fig. \ref{figure:twoTYPESpatient})
can be transformed into \xic{compact $\mathbf{SAD}$ representation (i.e., 10-dimensional feature vector, with entries listed in Eq.(\ref{eq:sad_dwpt}) and Eq.(\ref{eq:sad_dwt}) )}.
In Fig. \ref{figure:SADla012la038P}, we visualise compact $\mathbf{SAD}$ features corresponding to the paralysed sides of two patients (i.e., patients la012 and la040 from Fig.\ref{figure:twoTYPESpatient} ).
We \xic{notice} in the $\mathbf{SAD}$ feature space, it is not easy to distinguish the paralysed sides from these two different patients (in terms of CAHAI), indicating the necessity of developing more advanced stroke-related features (e.g., by also considering the non-paralysed side).

\subsubsection{Proposed Features}\hfill\\
Based on the compact $\mathbf{SAD}$ representation, we aim to further design effective features for reliable CAHAI score regression.
In Fig. \ref{figure:twoTYPESpatient} and Fig. \ref{figure:SADla012la038P}, we \xic{visualise} the behaviour patterns in different feature spaces.
Specifically, we \xic{plot} the \textbf{paralysed side} of patient la012 (with CAHAI score 55), and la040 (with CAHAI 26) using VM representation (Fig. \ref{figure:twoTYPESpatient}) and $\mathbf{SAD}$ representation (Fig. \ref{figure:SADla012la038P}).
From both figures, we can see the limitations of both representations.
Although VM can demonstrate distinct patterns from both patients, it may be also related to the large intra-class variability (e.g., personalised behaviour patterns). Moreover, the redundancy as well as the high-dimensionality make it hard for modelling.
On the other hand, \textbf{SAD} has low dimensionality, yet both patients exhibited high-level of similarity, indicating \xic{that} $\mathbf{SAD}$ of the paralysed side alone is not enough for distinguishing patients with different recovery levels.
\begin{figure}[H]
	\centering
	\includegraphics[width=5.5cm,height=4.2cm]{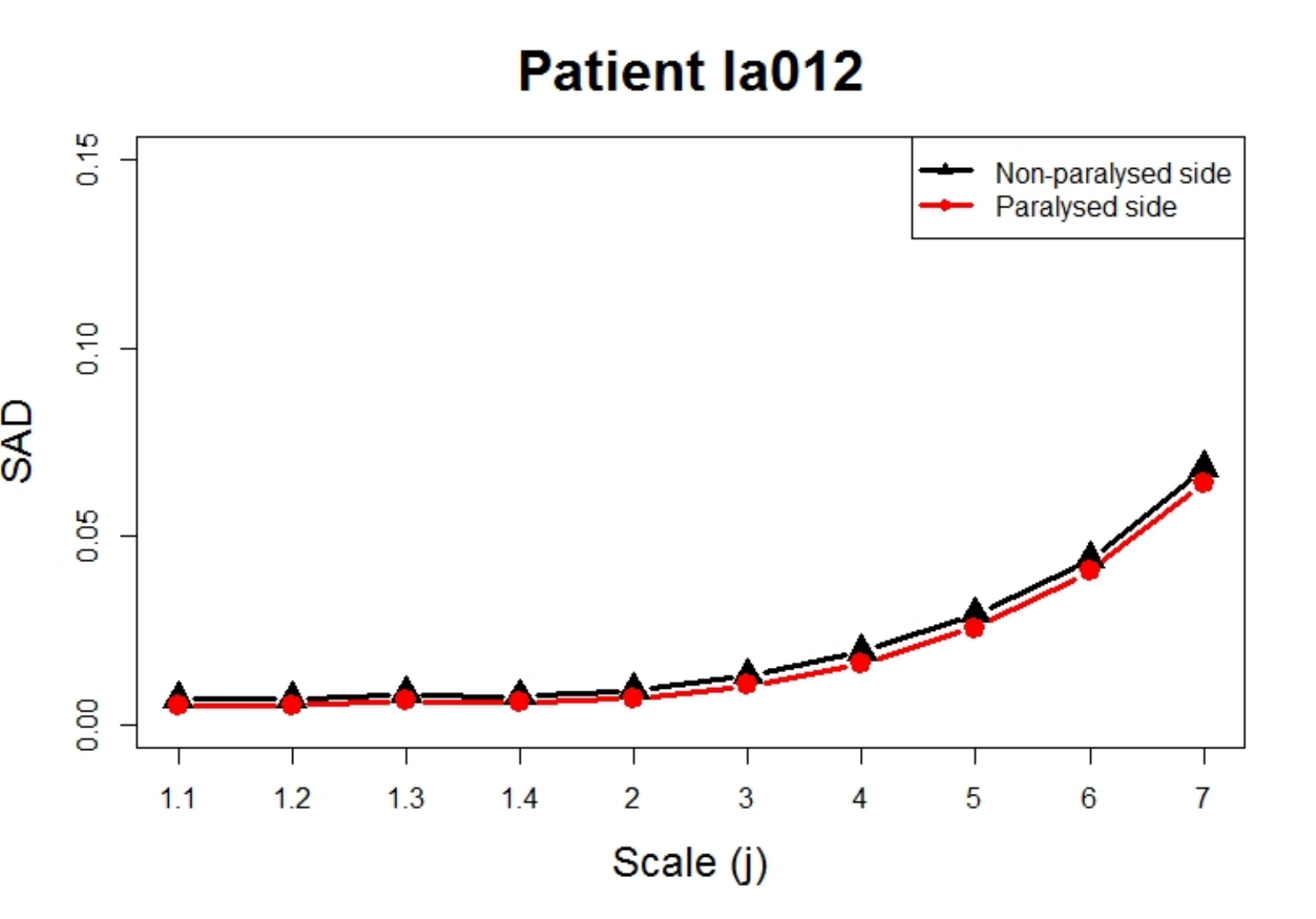}
	\includegraphics[width=5.5cm,height=4.2cm]{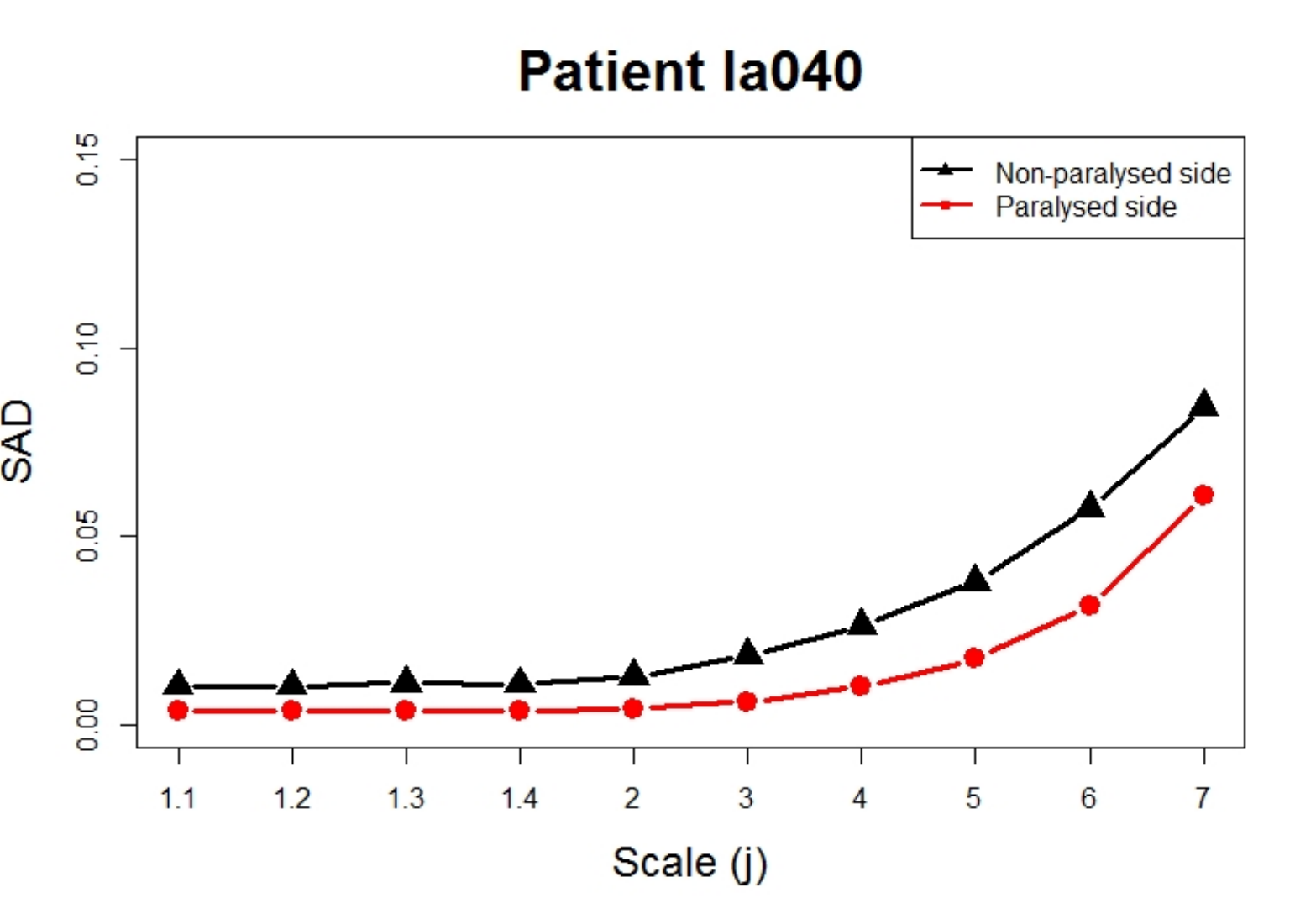}
	\caption{$\mathbf{SAD}$ representation with both paralysed/non-paralysed sides from two different patients (la012 with CAHAI score 55, and la040 CAHAI score 26). $\mathbf{SAD}$ features from the non-paralysed side may contain discriminant information for stroke-rehab modelling.}
	\label{figure:SADla027la038PNP}
\end{figure}
\begin{figure}[H]
	\centering
	\includegraphics[width=5.5cm,height=4.2cm]{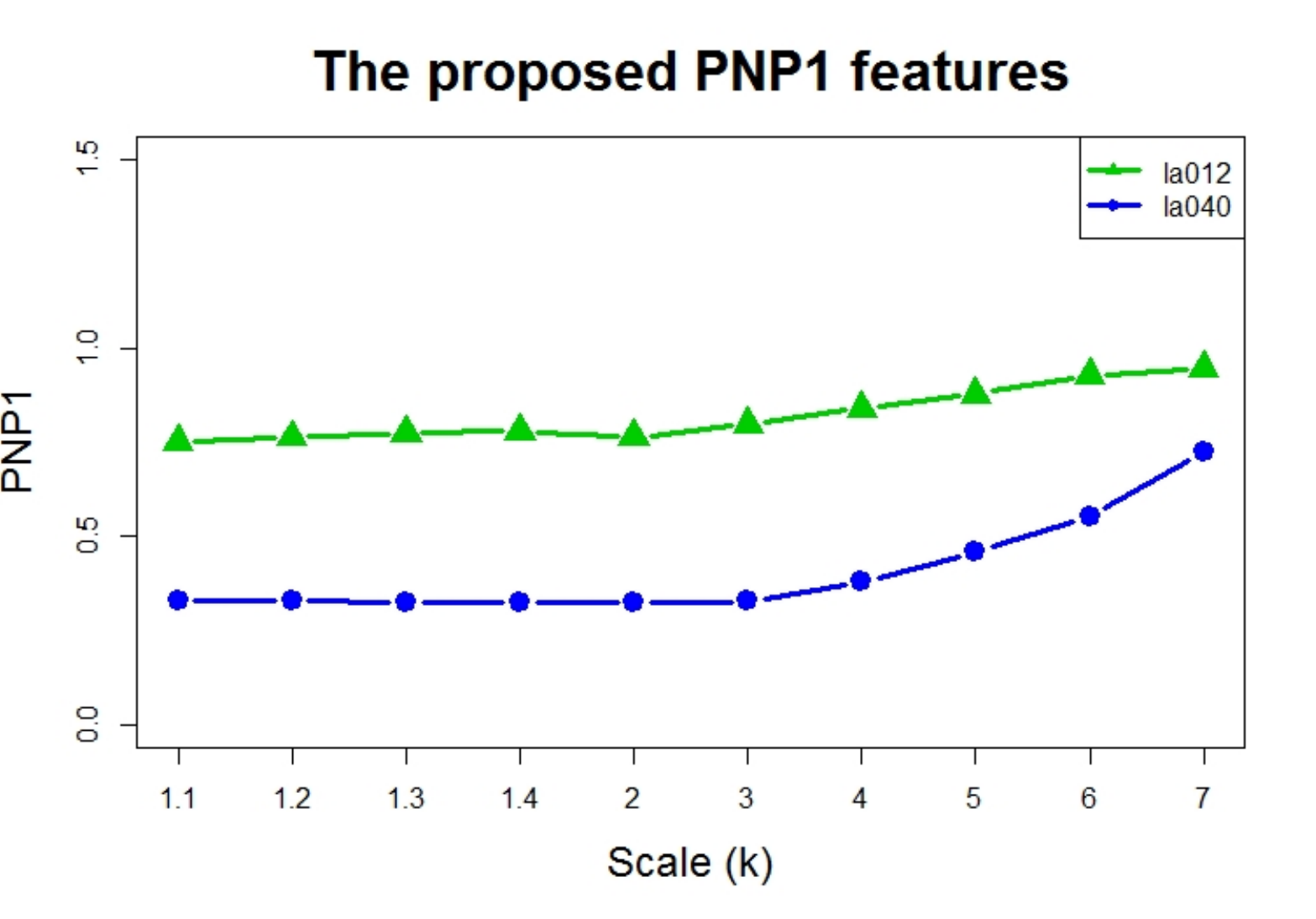}
	\includegraphics[width=5.5cm,height=4.2cm]{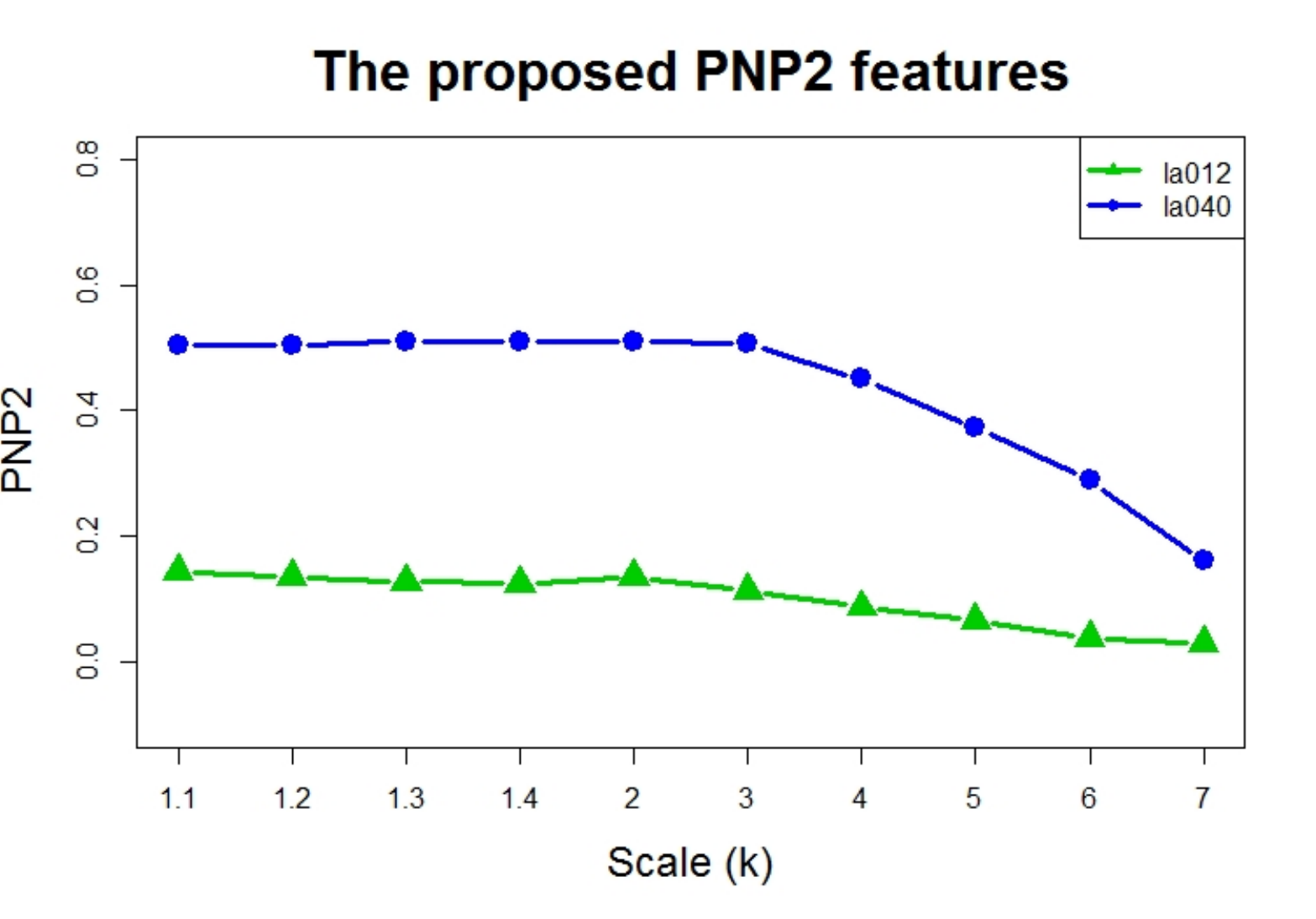}
	\caption{Two proposed $\mathbf{PNP}$ representations for two patients(la012, and la040), which can provide discriminant information in distinguishing the patients with different recovery levels (clinical CAHAI score)}.
	\label{figure:ratiola027la038PNP}
\end{figure}
Given the observations, we further \xic{visualise} \textbf{SAD} features from both paralysed/non-paralysed sides for both patients in Fig.\ref{figure:SADla027la038PNP}.
We can see patient la012 (with high recovery level) uses both hands (almost) equally while patient la040 (with low recovery level) tends to use the non-paralysed side more.
These observations \xic{motivate} us to design new features \xic{using} both sides, instead of the paralysed side alone.
\xic{In this work, we propose two types of features that combine both Paralysed side and Non-Paralysed side, namely 1) $\mathbf{PNP^1}$ that encodes the ratio information with entries defined as:
\begin{equation}\label{eq:pnp_1}
PNP^1_k = \frac{SAD_k^{p}}{SAD_k^{np}} 
\end{equation}
as well as its variant 2) $\mathbf{PNP^2}$ with entries defined as:
\begin{equation}\label{eq:pnp_2}
PNP^2_k = \frac{SAD_k^{np}-SAD_k^{p}}{SAD_k^{np}+SAD_k^{p}},
\end{equation}
where $k$ represents the scales defined in $\mathbf{SAD}$ features (as shown in Eq.(\ref{eq:sad_dwpt}) and Eq.(\ref{eq:sad_dwt}));}
$p$ and $np$ refer to the paralysed side and non-paralysed side respectively.
We also \xic{visualise} patient la012 and patient la040 using the new proposed features $\mathbf{PNP^1 }$ and $\mathbf{PNP^2 }$ in Fig. \ref{figure:ratiola027la038PNP}, from which we can see the proposed features can well distinguish these two patients, in contrast to \textbf{SAD} (Fig. \ref{figure:SADla012la038P}).
\xic{Although the proposed $\mathbf{PNP}$ features empirically exhibit the desired properties (i.e., compact and informative) for two patients, it should be pointed out that larger scale experiments should be conducted to evaluate the generalisation capability, which will be provided in the experimental section.}

\xic{
We summarise the procedure of generating $\mathbf{PNP}$ features as follows: 
\begin{enumerate}
\item Given 3-day raw accelerometer data, calculating the signal vector magnitude (VM) with the gravity effect removed;
\item calculating the second-wise VM (mean VM value for each second) as the new representation;
\item calculating \textbf{DWPT} features at scales $\{1.1, 1.2, 1.3, 1.4\}$ and \textbf{DWT} features at scales $\{2, 3, 4, 5, 6, 7\}$
\item given the \textbf{DWPT} and \textbf{DWT} features, calculating the 10-dimensional \textbf{SAD} features via Eq.(\ref{eq:sad_dwpt}) and Eq.(\ref{eq:sad_dwt}).
\item given \textbf{SAD} features, calculating the two proposed $\mathbf{PNP^1}$ and $\mathbf{PNP^2}$ features, via Eq.(\ref{eq:pnp_1}) and Eq.(\ref{eq:pnp_2}).
\end{enumerate}

}

\begin{table}
	\centering
	\begin{tabular}{|c|c|}
		\hline
		Feature type & \xic{F}eature entries for each type \\
		\hline
		$\mathbf{SAD^p}$   &  $SAD^p_{1.1}$, $SAD^p_{1.2}$, $SAD^p_{1.3}$,  $SAD^p_{1.4}$, $SAD^p_2$,  $SAD^p_3$, ... ,  $SAD^p_7$ \\
		\hline
		$\mathbf{SAD^{np}}$   &  $SAD^{np}_{1.1}$, $SAD^{np}_{1.2}$, $SAD^{np}_{1.3}$,  $SAD^{np}_{1.4}$, $SAD^{np}_2$, $SAD^{np}_3$, ... , $SAD^{np}_7$\\
		\hline
		$\mathbf{PNP^1 }$   &  $PNP^1_{1.1}$, $PNP^1_{1.2}$, $PNP^1_{1.3}$ ,$PNP^1_{1.4}$, $PNP^1_2$, $PNP^1_3$, ... , $PNP^1_7$\\
		\hline
		$\mathbf{PNP^2}$   &  $PNP^2_{1.1}$, $PNP^2_{1.2}$, $PNP^2_{1.3}$ ,$PNP^2_{1.4}$, $PNP^2_2$, $PNP^2_3$, ... , $PNP^2_7$ \\
		\hline
	\end{tabular}
	\caption{\xic{The proposed rehab-driven features}}
	\label{table:SADratios}
\end{table}

We \xic{list} 4 types of features, i.e., the original wavelet features extracted from paralysed ($\mathbf{SAD^p}$) and non-paralysed sides ($\mathbf{SAD^{np}}$) separately, as well as the two new proposed features ($\mathbf{PNP^1 }$ and $\mathbf{PNP^2 }$).
Based on 10 scales, we can form 40-dimensional feature vector, as shown in Table \ref{table:SADratios}.
However, there exist certain level of noises and redundancy (especially on $\mathbf{SAD^p}$, and $\mathbf{SAD^{np}}$), so it is crucial to develop feature selection mechanism or powerful prediction models for higher performance.

\subsection{Predictive models}
Based on the proposed representation, we aim to develop predictive models that can map features to the CAHAI score.
Although we \xic{reduce} the data redundancy significantly, there still exist data noises, which may encode irrelevant information.
It is crucial to develop robust mechanism to select the most relevant features, and here we \xic{use} a popular feature selection linear model (LASSO).
To model the nonlinear random effects in the longitudinal study, we also \xic{propose} to use the longitudinal mixed-effects model with Gaussian Process
prior (LMGP).
	
It is worth noting that our model will also take advantage of the medical history information (i.e., CAHAI score during the first visit) to predict CAHAI scores for the rest 7 weeks (i.e., week 2 - week 8).
From the perspective of practical application, CAHAI score from the initial week (referred to as $ini$) may be used as 
an important normalisation factor for different individuals.

	

\subsubsection{The linear fixed-effects model } \hfill\\
\label{section:LinearPredictvieModel}
Since there may exist some redundant or irrelevant features for the prediction task, first we \xic{propose} to use LASSO (Least Absolute Shrinkage and Selection Operator) for feature selection. 
	
Given the $41$-dimensional input variables (40 wavelet features \xic{listed in Table \ref{table:SADratios} and one} CAHAI score from the initial week), first we \xic{standardise} the data using z-norm, and each feature entry $x_k$ will be normalised as $x_k^{new} = (x_k-\overline{x}) / s_k$,
where $\overline{x}$ and $s_k$ are the mean and standard deviation of the $k^{th}$ feature.
Based on the aforementioned model, namely LASSO,  useful features can be selected, based on which prediction model can be developed.
For simplicity, we first \xic{use} linear model to predict the target CAHAI score ${y_i}$:
\begin{equation}
{y_{ij}} = \ve{x}_{ij}^\mathrm{T} \ve{\beta} + \epsilon_{ij},  \ \epsilon_{ij}\sim N(0, \sigma^2),
\label{equation:fixedeffects}
\end{equation}
where $i$ stands for the $i^{th}$ trial/visit (during week 2 - week 8) and $j$ represents the $j^{th}$ patients; $\ve{x}_{ij}$ represents the selected feature vector; $\ve{\beta}$ are the model parameter vector to be estimated, and $\epsilon_{ij}$ is the random noise term.
	
\subsubsection{Longitudinal mixed-effects model with Gaussian process prior (LMGP)} \hfill\\
\label{section:longitudinal}
It is simple to use linear model for CAHAI score prediction.
However, it ignores the heterogeneity nature among subjects in this longitudinal study.
To model the heterogeneity, we \xic{propose} to use a nonlinear mixed-effects model \cite{Shi:2012}, which consists of the fixed-effects part and random-effects part.
Specifically, the random-effects part contributes mainly on modelling the heterogeneity, making the the prediction process subject/time-adaptive for longitudinal studies.
The longitudinal mixed-effects model with Gaussian Process prior (LMGP) is defined as follows:
\begin{equation}
{y_{i,j}} = \ve{x}_{ij}^\mathrm{T} \ve{\beta} +  g(\ve{\phi}_{ij})
+ \epsilon_{ij}, \ \epsilon_{ij}\sim N(0, \sigma^2),
\label{equation:mixedeffects}
\end{equation}
where $i$,$j$ stand for the $i^{th}$ patient at the $j^{th}$ visit (from week 2 to week 8);
$\epsilon_{ij}$ refers to as independent random error and $\sigma^2$ is its variance;
In Eq\eqref{equation:mixedeffects}, $\ve{x}_{ij}^T\ve{\beta}$ is the fixed-effects part and $g(\ve{\phi}_{ij})$  represents the nonlinear random-effects part, and the latter can be modelled using 
a non-parametric Bayesian approach with a GP prior \cite{Shi:2012}.

It is worth noting that in LMGP the fixed-effects part $\ve{x}_{ij}^\mathrm{T}\ve{\beta}$ explains a linear relationship between
input features
and CAHAI, while the random-effects part $g(\ve{\phi}_{ij})$ is used to explain the variability caused by differences among individuals or time slots during different weeks.
By considering both parts, LMGP provides a solution of personalised modelling for this longitudinal data analysis.
In LMGP, it is important to select input features to model both parts, and we refer them to as fixed-effects features and random-effects features, respectively.
The effect of the fixed-effects features will be studied in the experimental evaluation section.

For LMGP training, we first \xic{ignore} the random-effects part, and only \xic{optimise} the parameters $\hat{\ve{\beta }}$ of the fixed-effects part (via ordinary least squares OLS);
With estimated parameters $\hat{\ve{\beta }}$, the residual $r_{ij}=y_{ij} - \ve{x}_{ij}^\mathrm{T} \hat{ \ve{\beta} } = g(\ve{\phi}_{ij}) + \epsilon_{ij}$ can be calculated, from which we can model the random-effects $$g(\ve{\phi}_{i,j})\sim GP(0,K(\cdot, \cdot ;\ve{\theta}) ).$$
In this paper we choose $K(\cdot, \cdot; \ve{\theta})$ 
as the following three different kernels (linear, squared exponential and rational quadratic), and here we take the squared exponential as an example.
The squared exponential (covariance) kernel function is defined as :
$
K\left(\ve \phi, \ve \phi' ; \ve{\theta}\right)=v_{0} \exp \left\{- d(\ve \phi, \ve \phi')/2 \right\} $ where $ d(\ve \phi, \ve \phi')=\sum_{q=1}^{Q} w_{q}\left({\phi_{i,j,q}}-{\phi_{i,j,q}^{\prime}}\right)^{2}$ is an extended distance between $ \ve \phi$ and $ \ve \phi'$. It involves the hyper-parameters $\ve{\theta} = (v_0, w_1, ... , w_Q )$. In Bayesian approach, we may choose the value of those parameters based on prior knowledge. It is however a difficult task due to the large dimension of $\ve \theta$. We used an empirical Bayesian method.

The training procedure include two steps. (I) Estimate $\ve \beta$ and $\sigma $ in equation \eqref{equation:fixedeffects}; (II) Estimate the values of the hyper-parameters $\ve \theta$ by an empirical Bayesian method, i.e. maximise the marginal likelihood from $\ve r_i \sim N(\ve 0, \vesub{C}{i}+\sigma^2 \ve I)$ for $i=1, \ldots, n$, where $\mathbf{C}_i\in \mathbb{R}^{J \times J}$ is the covariance matrix of $g(\cdot)$, and its element is defined by $K(\phi_{i,j}, \phi_{i,j'}; \ve \theta)$. To obtain a more accurate results, an iterative method may be used. Except the initial step, the error item in  \eqref{equation:fixedeffects} used in step I is replaced by
\[
\vesub{\epsilon}{i}=(\epsilon_1, \ldots, \epsilon_J) \sim N(\ve 0, \vesub{C}{i}+\sigma^2 \ve I))
\]
where all the parameters are evaluated by using the values obtained in the previous iteration.

The calculation of the prediction is relatively easy. The posterior distribution of $g(\vesub{\phi}{i})$ is a multivariate normal with mean $\mathbf{C}\left(\mathbf{C}+\sigma^{2} \mathbf{I}\right)^{-1} \ve r_{i}$ and the variance  $\sigma^{2} \mathbf{C} \left(\mathbf{C}+\sigma^{2} \mathbf{I}\right)^{-1} $.

The fitted value can therefore be calculated by the sum of $ \ve{x}^T_{ij} \hat{ \ve{\beta} }$ and the above posterior mean. The variance can be calculated accordingly. The detailed description can be found in \cite{shi:2011}.

\section{Experimental Evaluation}\label{section:experiment}
In this section, several experiments \xic{are} designed to evaluate the proposed features as well as the prediction systems.
The patients \xic{are} splitted into two groups according to the disease nature, i.e., the acute patient group (26 subjects) and the chronic patient group (33 subjects). \xic{Experiments are} conducted on both group separately.

Specifically for each group, leave one subject out cross validation(LOSO-CV) \xic{is} applied.
That is, for a certain group (acute or chronic) with $n$ subjects, in each iteration $1$ subject was used as test set while the rest $n-1$ subjects were used for training. This procedure \xic{is} repeated $n$ times to test all the $n$ subjects and average prediction performance (i.e., the mean predicted CAHAI) will be reported.

Since CAHAI score prediction is a typical regression problem, we \xic{use} the root mean square error (RMSE) as the evaluation metric, and lower mean RMSE values indicate better performance.


\subsection{Evaluation of the Proposed Feature $\mathbf{PNP}$}\label{section:featuresCOR}
In this subsection, we \xic{evaluate} the effectiveness of the proposed $\mathbf{PNP}$ features.
One most straight-forward approach is to calculate the correlation coefficients against the target CAHAI scores.
In Table \ref{table:corrCAHAI} we \xic{report} the corresponding correlation coefficients ($PNP^1_k$, and $PNP^2_k$ in 10 scales) for acute/chronic patients group.
The correlation coefficients of the original wavelet features (with paralysed side $SAD^p_k$, and non-paralysed side $SAD^{np}_k$ in 10 scales) against CAHAI score \xic{are} also reported for comparison.
From Table \ref{table:corrCAHAI}, we can see:
\begin{itemize}
\item $\mathbf{PNP}$ features generally have higher correlation coefficients (than $\mathbf{SAD}$) against the CAHAI scores.
\item for $\mathbf{PNP}$ features, from Scale $k=1.1$ to $k=5$ there are higher correlations against the CAHAI scores.
\item  for chronic patients, $\mathbf{SAD}$ features (on the non-paralysed side) exhibit comparable correlation scores with $\mathbf{PNP}$ features.
\end{itemize}
These observations indicate the necessities of selecting useful features on building the prediction system.
Although $\mathbf{PNP}$ demonstrates more powerful prediction capacity, in some cases, $\mathbf{SAD}$ (e.g., extracted from the non-paralysed side) may also provide important information for a certain population (e.g., chronic patients).

\begin{table}[ht]
	\centering
	\begin{tabular}{|c|cccc|cccc|}
		\hline
		-	&  &Acute &Patients & 	&  &Chronic &Patients & \\
		\hline
		Scale (k)	& $SAD^p_k$ &  $SAD^{np}_k$ & $PNP^1_k$ & $PNP^2_k$ & $SAD^p_k$ &  $SAD^{np}_k$ & $PNP^1_k$ & $PNP^2_k$ \\
		\hline
		k=1.1 & -0.41 & 0.32 & 0.68 & -0.70 & 0.22 & 0.49 & 0.56 & -0.56\\
		k=1.2 & -0.42 & 0.33 & 0.69 & -0.71 & 0.24 & 0.50 & 0.57 & -0.56\\
		k=1.3 & -0.43 & 0.32 & 0.70 & -0.72 & 0.23 & 0.51 & 0.58 & -0.57\\
		k=1.4 & -0.42 & 0.33 & 0.69 & -0.71 & 0.24 & 0.51 & 0.57  & -0.57\\
		k=2   & -0.42 & 0.31 & 0.69 & -0.71 & 0.23 & 0.50 & 0.56  & -0.55\\
		k=3   & -0.42 & 0.27 & 0.67 & -0.68 & 0.25 & 0.50 & 0.53  & -0.52\\
		k=4   & -0.43 & 0.20 & 0.60 & -0.63 & 0.26 & 0.50 & 0.48  & -0.47\\
		k=5   & -0.42 & 0.10 & 0.49 & -0.52 & 0.27 & 0.50 & 0.43  & -0.42\\
		k=6   & -0.37 & -0.01 & 0.35 & -0.38 & 0.27 & 0.48 & 0.35  & -0.34\\
		k=7   & -0.30 & -0.10 & 0.19 & -0.20 & 0.28 & 0.45 & 0.25  & -0.24\\
		\hline
	\end{tabular}
	\caption{Correlation coefficients of the wavelet features and CAHAI score.}
	\label{table:corrCAHAI}
\end{table}



\begin{figure} \centering
	\subfigure { \label{fig:a}
		\includegraphics[width=13cm,height=8cm]{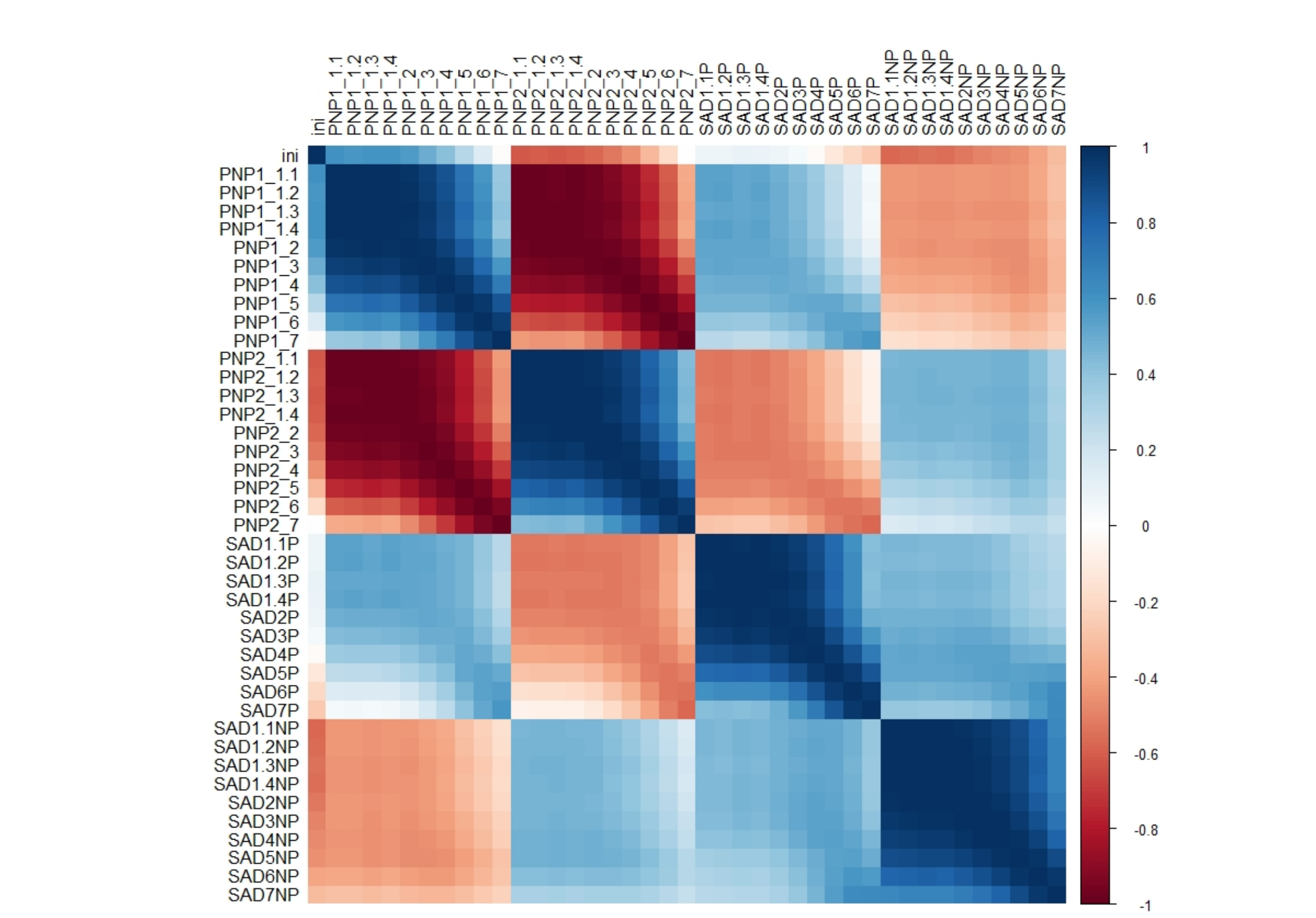} }
	\subfigure { \label{fig:b}
		\includegraphics[width=13cm,height=8cm]{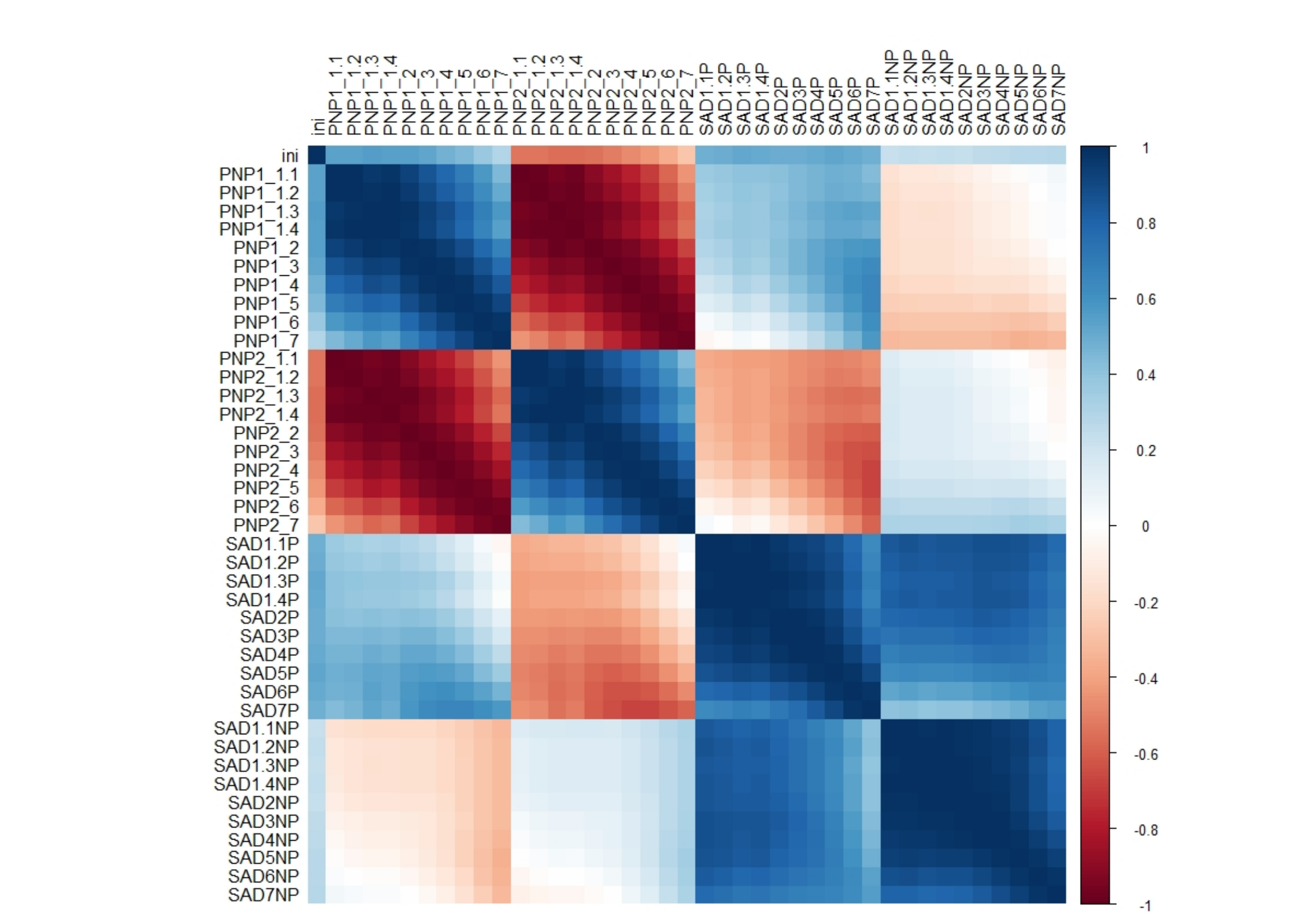} }
	\caption{Cross-correlation of the candidate features for two patient groups (top: acute patients; bottom: chronic patients).
 \xic{In general, $\mathbf{PNP}$ features, $\mathbf{SAD}$ features and the medical history information $ini$ are less correlated, compared with within-feature correlation (e.g., within $\mathbf{PNP}$ features ) }
}
	\label{figure:correlationacutechronic}
\end{figure}
For better understanding the relationship between these features, we also \xic{report} the cross-correlation between each feature pairs.
Noting we also \xic{include} the medical history feature, i.e., the initial week-1 CAHAI score.
From Fig. \ref{figure:correlationacutechronic}, and we have the following observations:
\begin{itemize}
\item For both patient groups, the $\mathbf{PNP}$ features are highly correlated. $\mathbf{PNP}$ features within the same type ($\mathbf{PNP}^1$ or $\mathbf{PNP}^2$) tend to be positively correlated, while $\mathbf{PNP}$ features from different types tend to be negatively correlated.
\item For acute patients, $\mathbf{SAD}$ features for each side (paralysed side $\mathbf{SAD^p}$ or non-paralysed side $\mathbf{SAD^{np}}$) are highly (positively) correlated, yet the $\mathbf{SAD}$ features from different sides are less correlated. For chronic patients, however, $\mathbf{SAD}$ features from both sides are highly (positively) correlated.
\item In general, $\mathbf{PNP}$ features, $\mathbf{SAD}$ features and the medical history information $ini$ are less correlated, indicating them as potentially complementary information to be fused.
\end{itemize}
Based on the above findings, it is clear that within each feature types, there may exist high-level of feature redundancy, and it is necessary to select the most relevant feature subsets.
For acute and chronic patient groups, the optimal feature subset may vary due to the different movement patterns (e.g., on paralysed/non-paralysed sides).
Although the proposed $\mathbf{PNP}$ features can alleviate this problem to some extent, it is beneficial to combine the less correlated features (i.e.,$\mathbf{PNP}$, $\mathbf{SAD}$, and $ini$).

\subsection{Evaluation of the Predictive Models}\label{section:predictiveresults}

\subsubsection{Feature Selection} \hfill\\
Based on the feature correlation analysis in Sec. \ref{section:featuresCOR}, it is important we select the most relevant features from various sources (i.e., $\mathbf{PNP}$, $\mathbf{SAD}$, and $ini$).
Different from the correlation-based approach which can select each feature independently (by the correlation coefficient), LASSO can select the features by solving a linear optimisation problem with sparsity constraint, and it takes the relationship of the features into consideration.
Based on LASSO we \xic{select} the most important features for both acute/chronic patients, as shown in Table \ref{table:variableselection}.
\begin{table}[ht]
		\newcommand{\tabincell}[2]{\begin{tabular}{@{}#1@{}}#2\end{tabular}}
		\centering
		\begin{tabular}{|c|c|}
			\hline
			\tabincell{c}{Acute Patients}   & \multicolumn{1}{|c|}{Chronic Patients} \\
			\hline
			\tabincell{l}{$PNP^2_3$, $PNP^1_{6}$, $SAD_2^{np}$, $SAD_{1.2}^{p}$\\ $SAD_6^{np}$, $ini$}  & \tabincell{l}{$PNP^1_{1.4}$, $SAD_4^{p}$, $SAD_2^{np}$, $PNP^2_{1.3}$\\ $PNP^1_4$,  $PNP^2_{1.1}$, $ini$, $PNP^1_6$\\ $SAD_{1.4}^{np}$, $SAD_6^{np}$ }\\
			\hline
		\end{tabular}
		\caption{Selected features using LASSO
		}
		\label{table:variableselection}
\end{table}


It is also worth mentioning that the
wavelet-based features can bring certain levels of interpretability.
$SAD_j$ represents the point energy in the signal at the decomposition level $j$ based on the energy preserving condition (see Appendix \ref{section:waveletfeatures} for more details).
Specifically, it relates to the degree of energy among the different activity levels (in different frequency domain based on the decomposition scale $j$).
The activities such as  jumping or lifting an object may correspond to high-frequency signal, while sedentary or eating may be low-frequency signal.
Based on these, we can interpret the key features in Table \ref{table:variableselection}.
For example, for acute patients key features (which is high-related to stroke-rehab modelling) correspond to asymmetric activities in low/medium-frequency level (i.e., with $PNP_3^2, PNP_6^1$), non-paralysed-based activities in low/medium-frequency level(i.e., with $SAD_2^{np}, SAD_6^{np}$), and paralysed-side based activities in high-frequency level (i.e.,with $SAD_{1.2}^p$).

\subsubsection{Performance of linear fixed-effects model}
\begin{figure}[H]
	\centering
	\includegraphics[width=5.5cm,height=4.2cm]{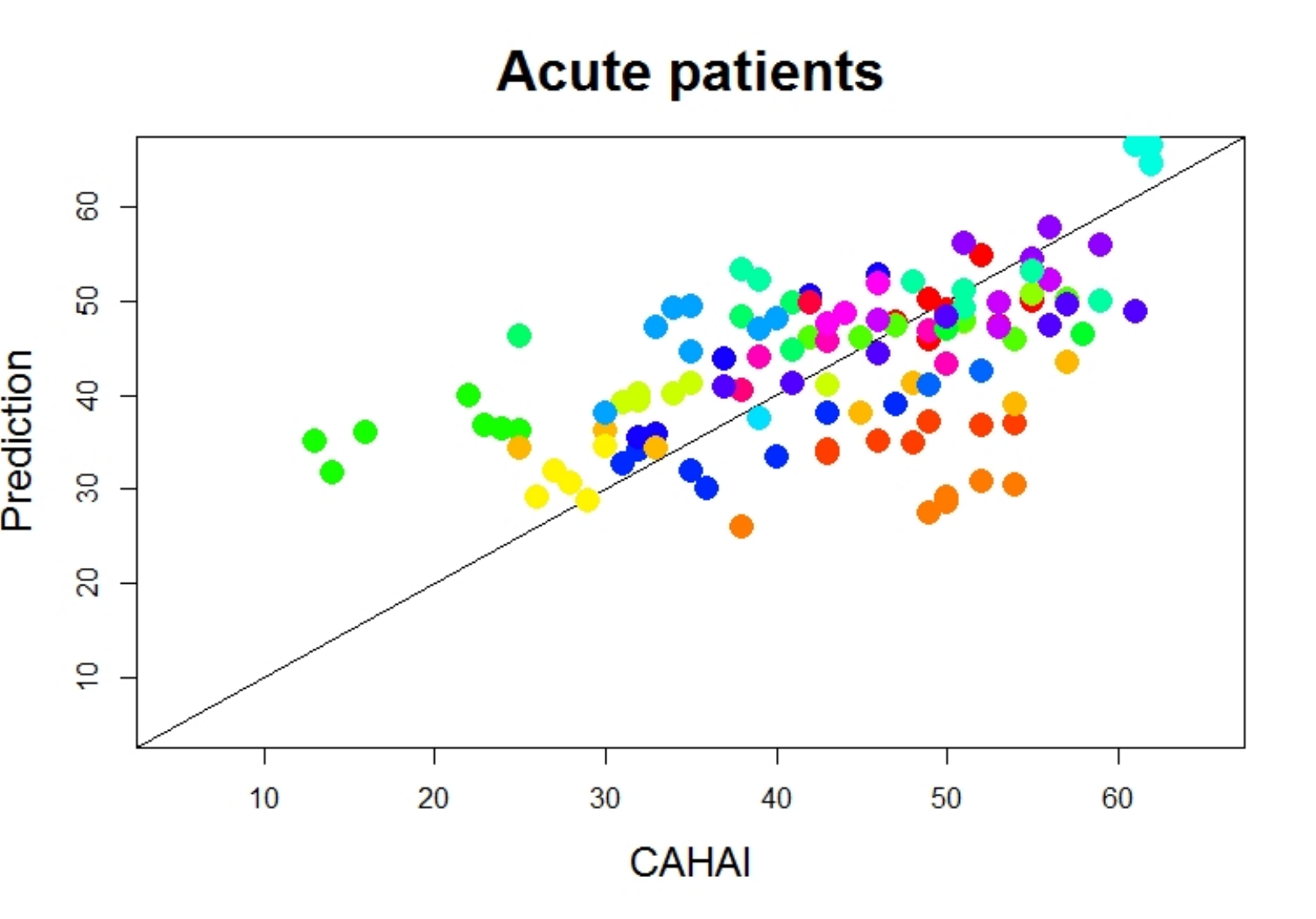}
	\includegraphics[width=5.5cm,height=4.2cm]{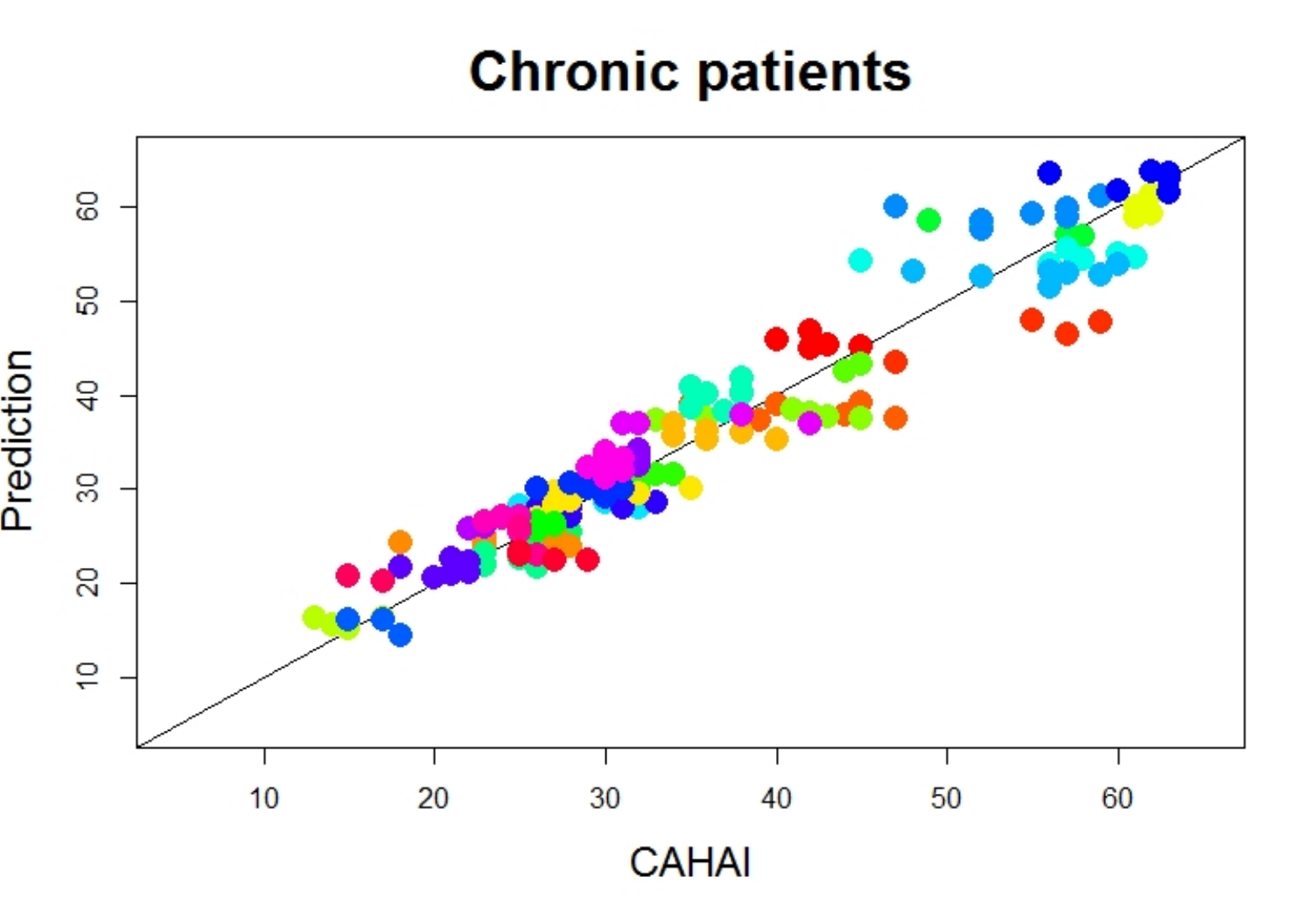}
	\caption{Linear model prediction vs clinical CAHAI; Left: Acute patients (RMSE 7.24); Right: Chronic patients (RMSE 3.29).
Each point corresponds to a trial (i.e., data collected from 3 days), and different colours represent different subjects.
	}
	\label{figure:LM}
\end{figure}

\begin{figure}[H]
	\centering
	\includegraphics[width=5.5cm,height=4.2cm]{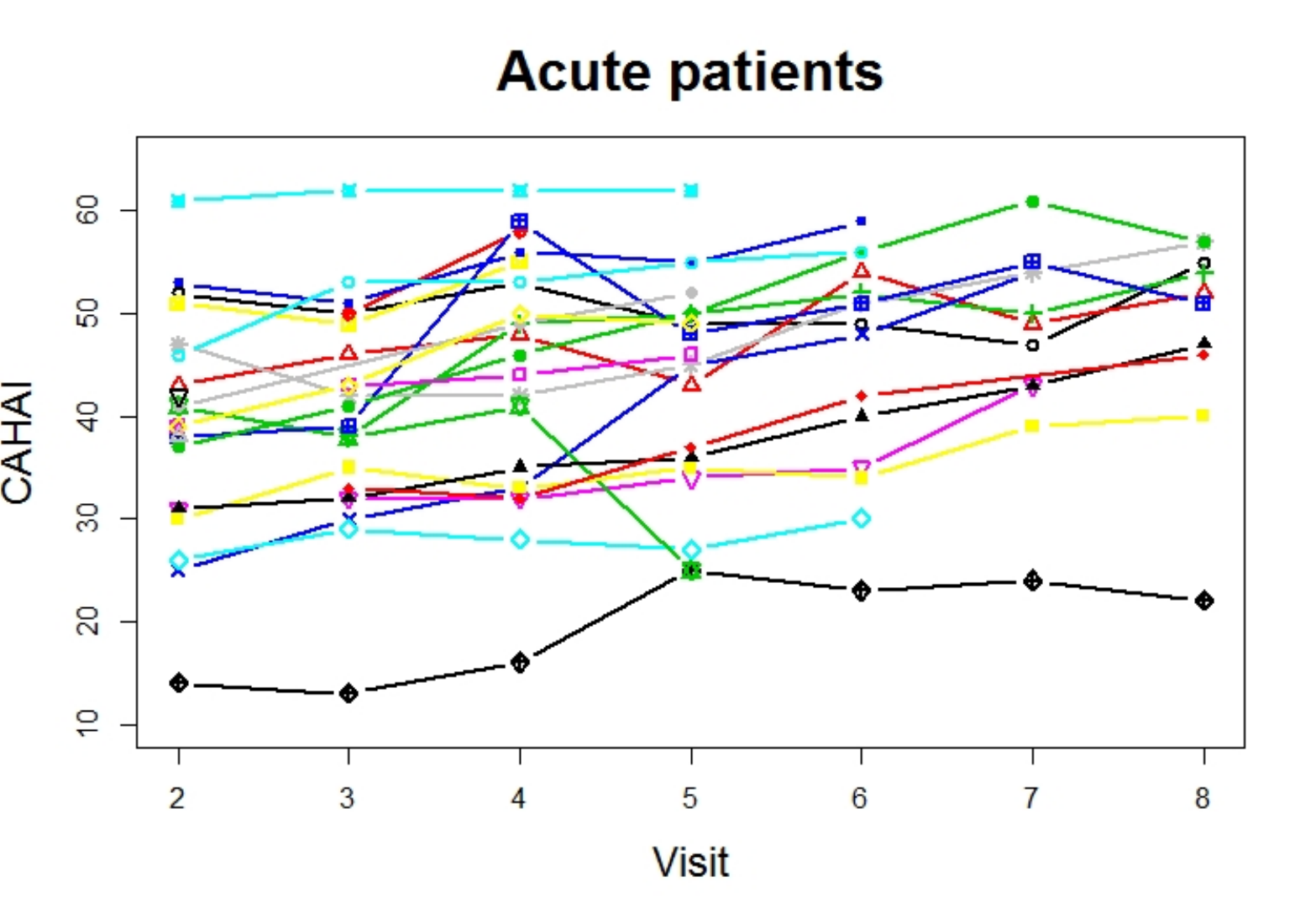}
	\includegraphics[width=5.5cm,height=4.2cm]{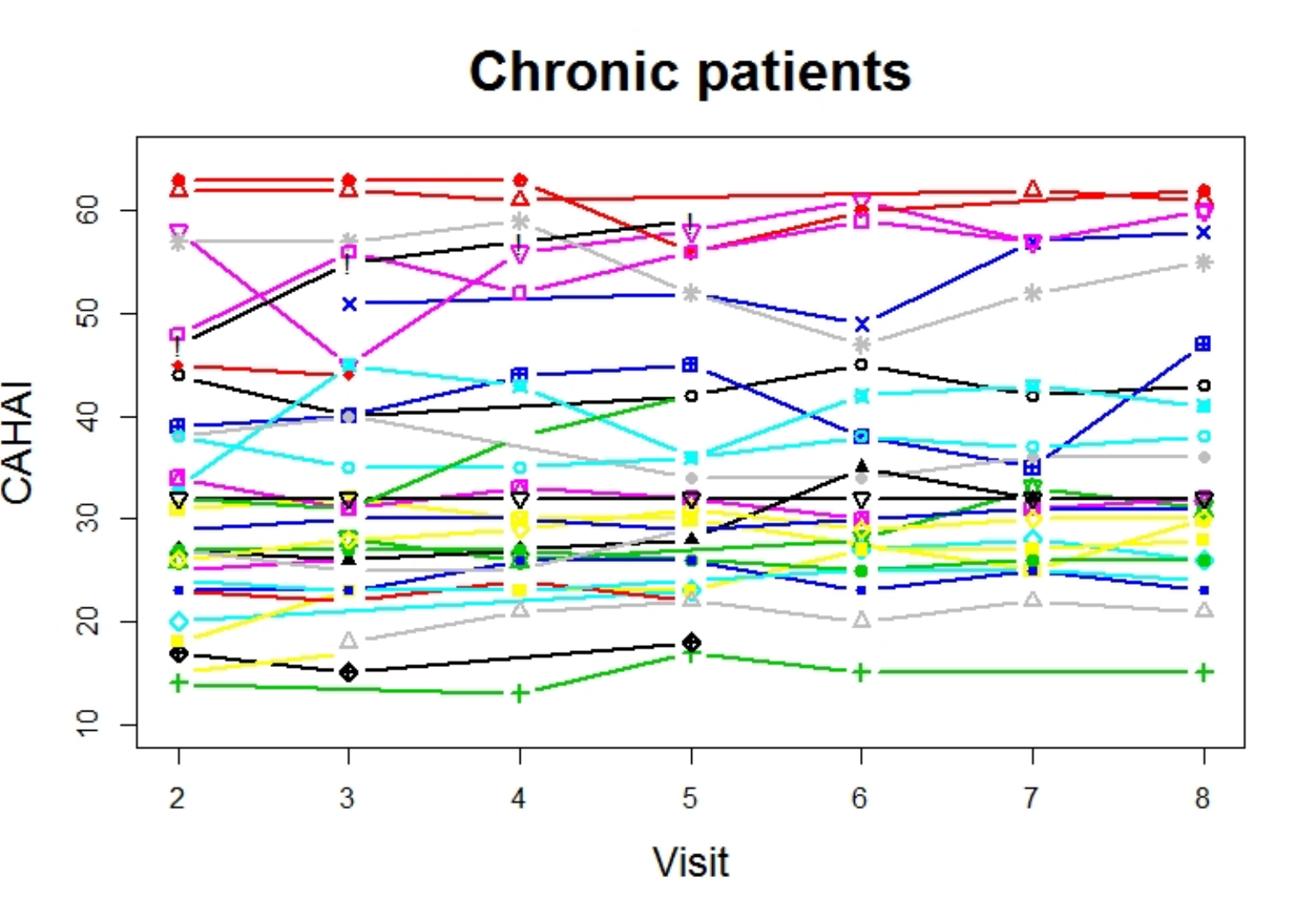}
	\caption{Clinical assessed CAHAI distribution with respect to visit; Stroke rehabilitation levels may be stable for chronic patient while may vary substantially for acute patients.
	}
	\label{figure:cahailongitudinal}
\end{figure}

\hfill\\
Based on the selected features, we \xic{perform} leave-one-patient-out cross validation on these two patient groups respectively using the linear fixed-effects model.
As shown in Fig. \ref{figure:LM}, the prediction results of the chronic patients (with mean RMSE 3.29) tend to be much better than the ones of the acute group (with mean RMSE 7.24).
One of the main reasons might be the nature of the patient group.
In Fig. \ref{figure:cahailongitudinal}, we \xic{plot} the clinical CAHAI distribution (i.e., the ground truth CAHAI) from week 2 to week 8, and we can see the clinical CAHAI scores are very stable for chronic patients.
On the other hand, for acute patients who suffered from stroke in the past 6 months, their health statuses were less stable and affected significantly by various factors, and in this case the simple linear fixed-effected model yields less promising results.


\subsubsection{Performance of Longitudinal mixed-effects Model with Gaussian Process prior (LMGP)} \hfill\\
We also \xic{develop} LMGP for both patient groups.
We have applied different covariance kernels in LMGP models and found the one with powered exponential kernel achieves the best results. The following discussion will therefore focus on the model with this kernel. More results of using other kernels can be found in Appendix. \ref{section:LMGP_kernels}.
\begin{figure}[H]
	\centering
	\includegraphics[width=5.5cm,height=4.2cm]{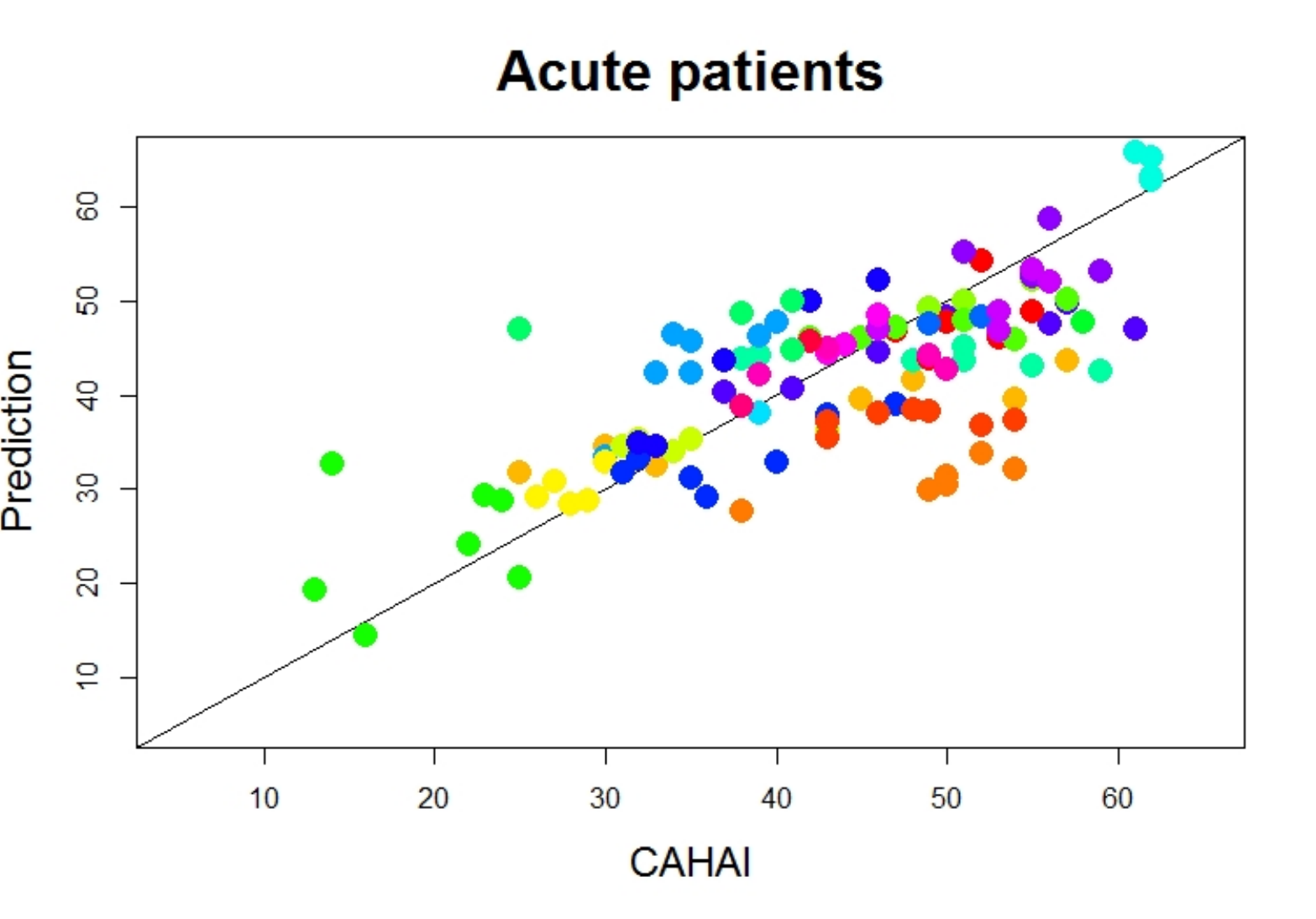}
	\includegraphics[width=5.5cm,height=4.2cm]{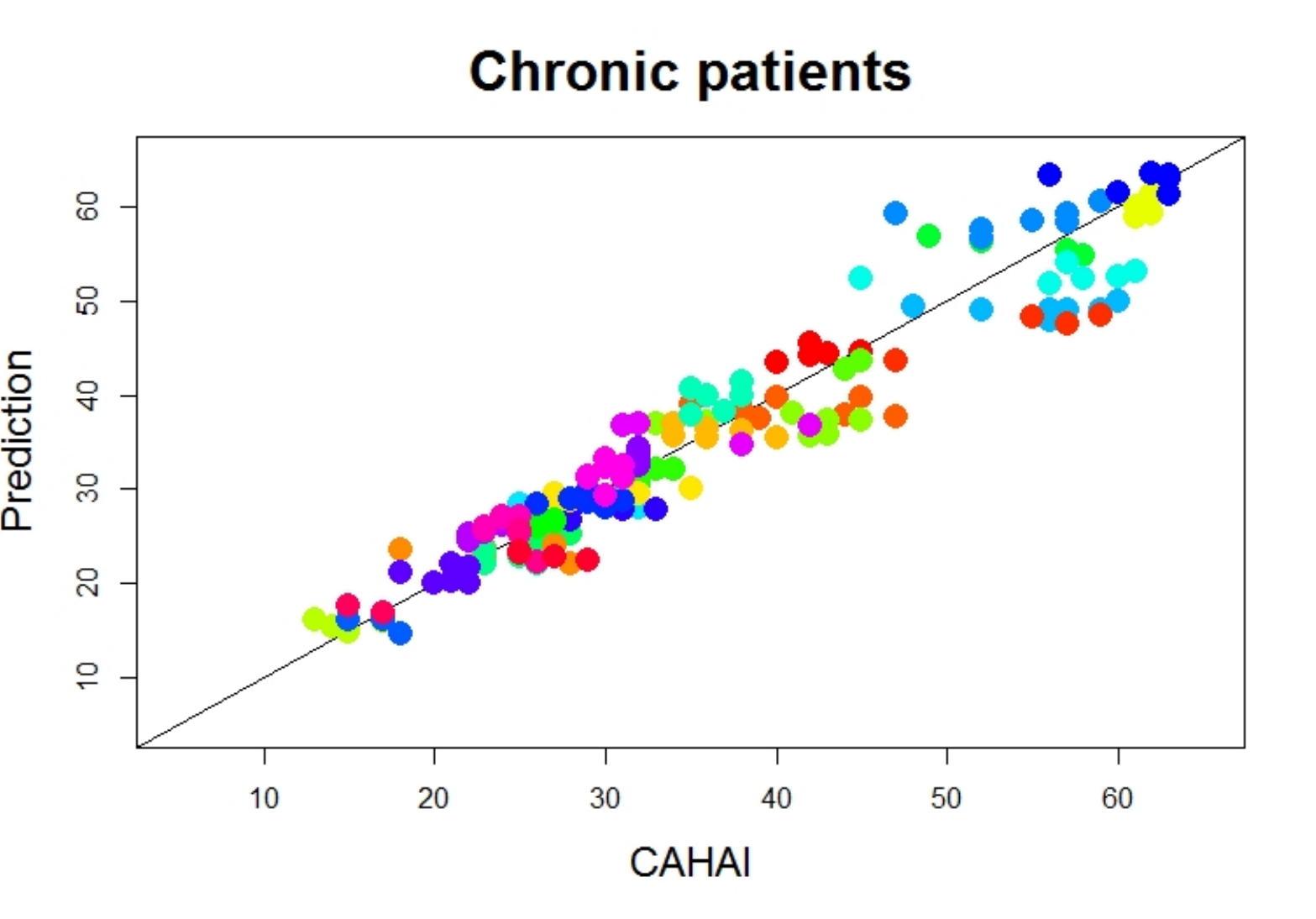}
	\caption{LMGP prediction vs clinical CAHAI; Left: Acute patients (RMSE 5.75); Right: Chronic patients (RMSE 3.12).
Each point corresponds to a trial (i.e., data collected from 3 days), and different colours represent different subjects.}
	\label{figure:LMGP}
\end{figure}
Here, we \xic{use} the selected features (from Table \ref{table:variableselection}) as the fixed-effects features and random-effects features.
Similar to the linear fixed-effects model, we \xic{evaluate} the performance based on leave-one-patient-out cross validation, and the mean RMSE values \xic{are} reported in 
Fig. \ref{figure:LMGP}, from which can see LMGP can further reduce the errors when compared with the fixed-effects linear model, with mean RMSE 5.75 for acute patients and 3.12 for chronic patients, respectively.
\begin{figure}[H]
	\centering
	\includegraphics[width=5.5cm,height=3.5cm]{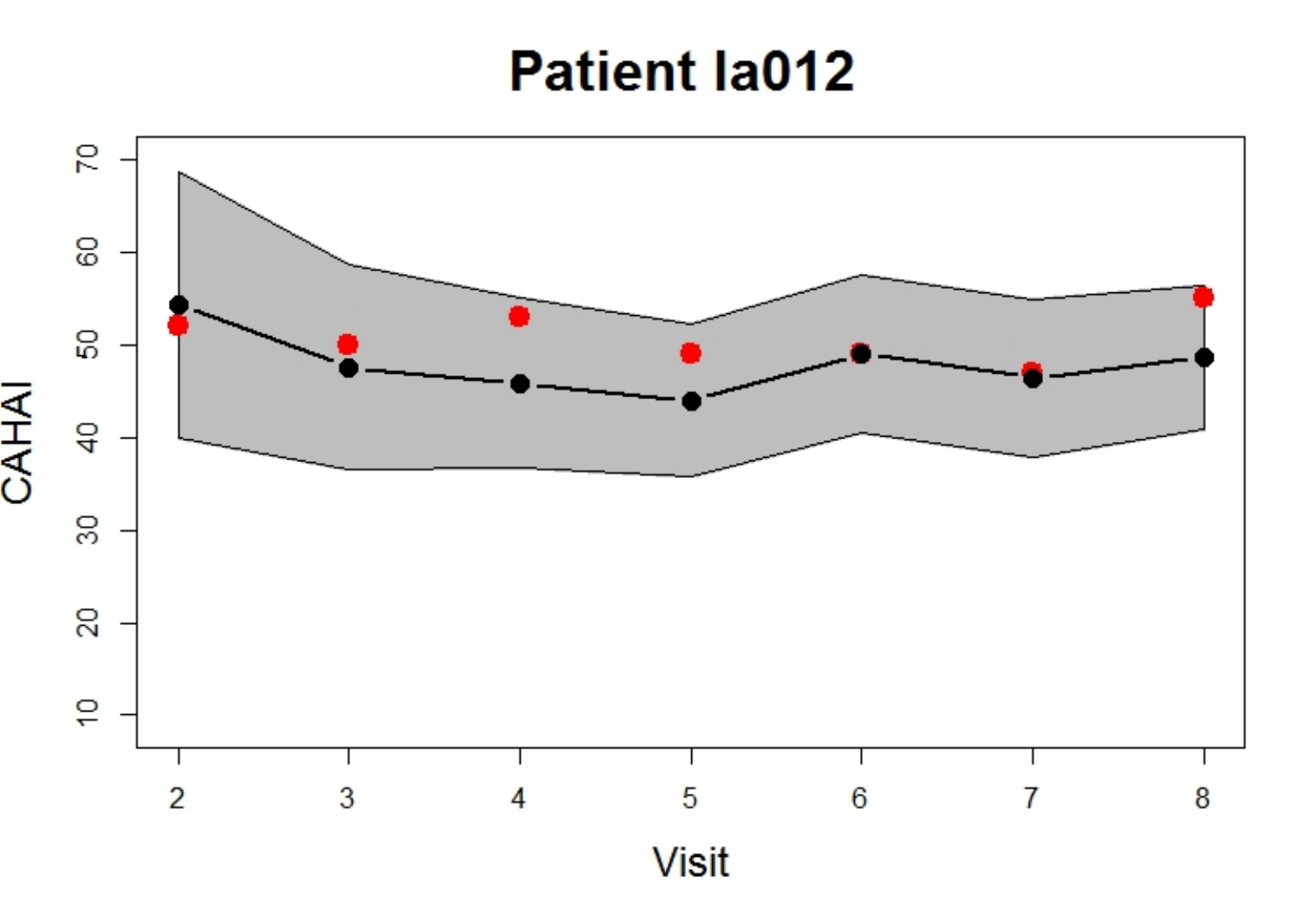}
	\includegraphics[width=5.5cm,height=3.5cm]{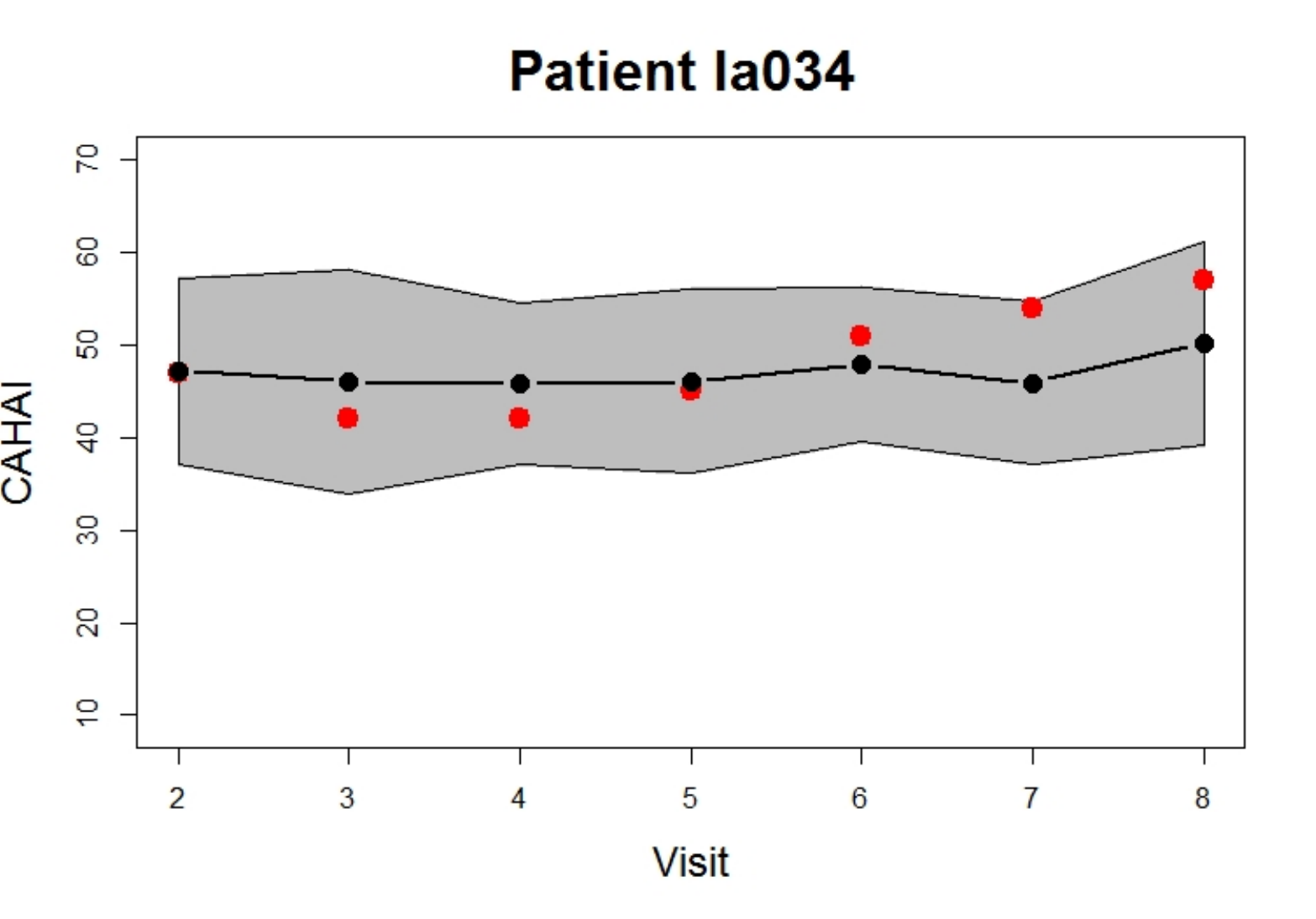}
	\includegraphics[width=5.5cm,height=3.5cm]{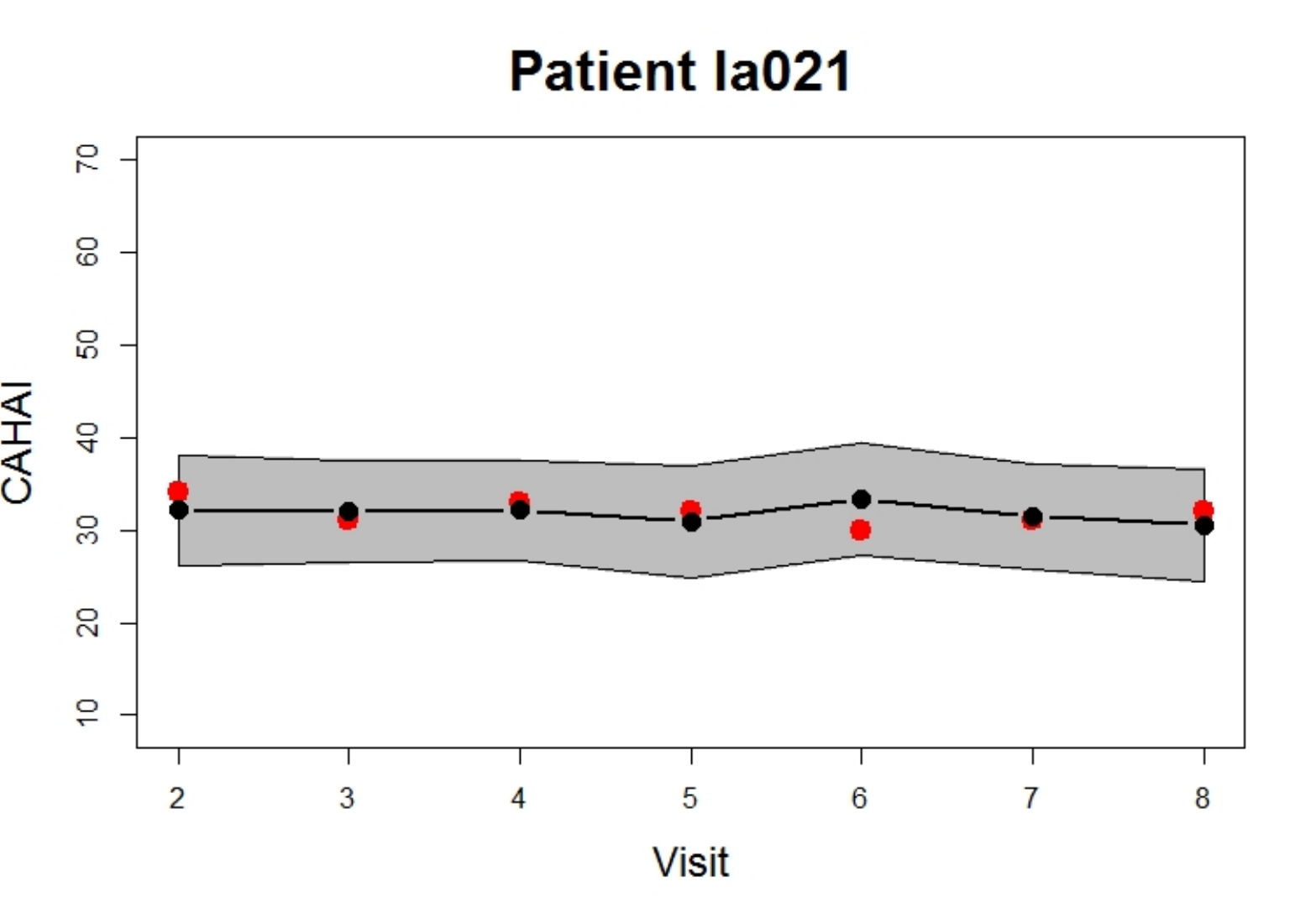}
	\includegraphics[width=5.5cm,height=3.5cm]{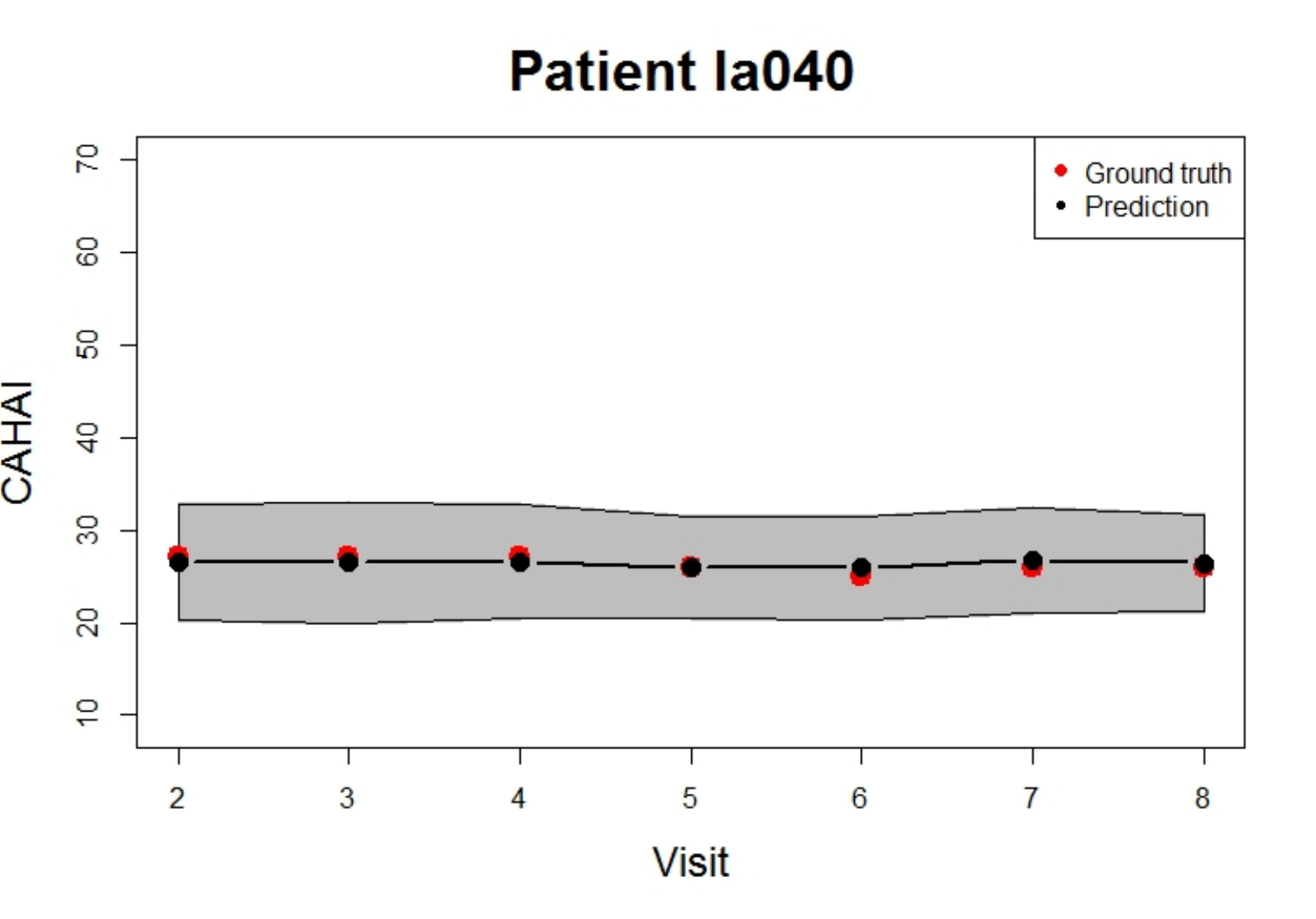}
	\caption{Continues monitoring using LMGP for 4 patients (top: two chronic patients; bottom: two acute patients); Dark points are the trial-wise/week-wise (i.e., each trial including data collected from 3 days per week) prediction and red points are the corresponding ground truth CAHAI scores.
	}
	\label{figure:95piforacutechronic}
\end{figure}
Based on LMGP, we also \xic{perform} "continuous monitoring"---with week-wise predicted CAHAI score --- on 4 patients (two for each patient group) from week 2 to week 8, and the results \xic{are} reported (with mean and $95\%$ confidence interval) in Fig. \ref{figure:95piforacutechronic}, which is extremely helpful when uncertainty measurement is required.

\subsubsection{On the fixed-effects part of LMGP} \hfill\\
LMGP includes two key parts, i.e., the linear fixed-effects and the non-linear random-effects part, and it is important to choose the key features for modelling.
Since the fixed-effects part measures the main (linear) relationship between the input features and the predicted CAHAI, we \xic{study} the corresponding feature subsets.
For random-effects part, we \xic{use} the full LASSO features (as shown in Table \ref{table:variableselection}).

To select the most important feature subset for the fixed-effects part modelling, we \xic{rank} the features (from Table \ref{table:variableselection}) based on two criteria: LASSO coefficients, and correlation coefficients (between features and CAHAI, as described in Sec.\ref{section:featuresCOR}).
Table \ref{table:variableRanking} demonstrates ranked features, and here only the top $50\%$ features (i.e., top 3 features for acute patients and top 5 features for chronic patients) \xic{are} used to model the fixed-effects part, and the settings as well as the results \xic{are} reported in Table \ref{table:acutechronic3models}.


	

\begin{table}[ht]
		\newcommand{\tabincell}[2]{\begin{tabular}{@{}#1@{}}#2\end{tabular}}
		\centering
		\begin{tabular}{|c|c|c|}
			\hline
			- & \tabincell{c}{Acute Patients}   & \tabincell{c}{Chronic Patients} \\
			\hline
			\tabincell{c}{LASSO \\Coefficients\\ (absolute value)} & \tabincell{l}{$PNP^2_3$, $PNP^1_{6}$, $SAD_2^{np}$, $SAD_{1.2}^{p}$\\ $SAD_6^{np}$, $ini$}  & \tabincell{l}{$PNP^1_{1.4}$, $SAD_4^{p}$, $SAD_2^{np}$, $PNP^2_{1.3}$\\ $PNP^1_4$,  $PNP^2_{1.1}$, $ini$, $PNP^1_6$\\ $SAD_{1.4}^{np}$, $SAD_6^{np}$ }\\
			\hline
			\tabincell{c}{Correlation \\Coefficients\\ (absolute value)} & \tabincell{l}{$PNP^2_3$, $ini$, $SAD_{1.2}^{p}$, $PNP^1_{6}$\\ $SAD_2^{np}$, $SAD_6^{np}$}  & \tabincell{l}{$ini$, $PNP^1_{1.4}$, $PNP^2_{1.3}$, $PNP^2_{1.1}$\\ $SAD_{1.4}^{np}$, $SAD_2^{np}$,  $PNP^1_4$,  $SAD_6^{np}$\\$PNP^1_6$,  $SAD_4^{p}$}\\
			\hline
		\end{tabular}
		\caption{Feature importance ranking (based on two criteria) for acute/chronic patients.
		}
		\label{table:variableRanking}
\end{table}

It is interesting to observe the performance may \xic{vary when different feature subsets are applied.}
Specifically, with the top feature subsets, modelling the LMGP's fixed-effects part can further reduce the errors \xic{to some extent} for acute patients, in contrast to chronic patients with increased errors.
The top 5 features selected via the LASSO criterion yields the worst performance for chronic patients, and one possible explanation could be the lack of feature $ini$ -----the initial health condition-----a major attribute for chronic patient modelling (see Fig. \ref{figure:cahailongitudinal}).

\begin{table}[ht]

	\newcommand{\tabincell}[2]{\begin{tabular}{@{}#1@{}}#2\end{tabular}}
	\centering
	\begin{tabular}{|c|c|c|c|}
		\hline
		\multirow{4}*{\rotatebox{90}{\textbf{Acute Patients}}}
		&  \tabincell{c}{\textbf{Fixed-effects} \\ \textbf{features}  }   & \tabincell{c}{\textbf{Random-effects}\\ \textbf{features} } & \textbf{RMSE} \\
		\cline{2-4}
		 & \tabincell{c}{full 6 features in Table \ref{table:variableselection}}  & \tabincell{c}{full 6 features \\in Table \ref{table:variableselection}} & 5.75 \\
		\cline{2-4}
		  & \tabincell{c}{top 3 features (Corr criterion in Table \ref{table:variableRanking}):\\ $PNP^2_3$, $ini$, $SAD_{1.2}^{p}$}  & \tabincell{c}{full 6 features \\in Table \ref{table:variableselection}} & 5.37 \\
		
		\cline{2-4}
	      & \tabincell{l}{top 3 features (LASSO criterion in Table \ref{table:variableRanking}):\\$PNP^2_3$, $PNP^1_{6}$, $SAD_2^{np}$}  & \tabincell{c}{full 6 features \\in Table \ref{table:variableselection}} & 5.51  \\
		\hline
		\hline
		\multirow{4}*{{\rotatebox{90}{\textbf{Chronic Patients}}}}
		&  \tabincell{c}{\textbf{Fixed-effects} \\ \textbf{features}  }   & \tabincell{c}{\textbf{Random-effects}\\ \textbf{features}  } & \textbf{RMSE} \\
		\cline{2-4}
		  & \tabincell{c}{full 10 features in Table \ref{table:variableselection}}  & \tabincell{c}{full 10 features \\in Table \ref{table:variableselection}} & 3.12 \\
		\cline{2-4}
		  & \tabincell{c}{top 5 features (Corr criterion in Table \ref{table:variableRanking}):\\$ini$, $PNP^1_{1.4}$, $PNP^2_{1.3}$ $PNP^2_{1.1}$, $SAD_{1.4}^{np}$}  & \tabincell{c}{full 10 features \\in Table \ref{table:variableselection}} & 3.20 \\		
		\cline{2-4}
		   & \tabincell{c}{top 5 features (LASSO criterion in Table \ref{table:variableRanking}):\\$PNP^1_{1.4}$, $SAD_4^{p}$, $SAD_2^{np}$ $PNP^2_{1.3}$, $PNP^1_4$}  & \tabincell{c}{full 10 features \\in Table \ref{table:variableselection}} & 5.12 \\
		\hline
	\end{tabular}
	\caption{LMGP's fixed-effects part modelling results (RMSE) based on different feature subsets
	}
	\label{table:acutechronic3models}
\end{table}

\subsubsection{Model comparison} \hfill\\

Based on our proposed (41-dimensional) stroke-rehab-driven features, we \xic{compare} LMGP with a number of classical predictive models, such as neural network (NN), support vector regression (SVR) and random forest regression(RF) for acute/chronic patient groups.
It is worth noting that we cannot use the popular deep learning structures such as convolutional neural network(CNN) or recurrent neural network(RNN) on the time-series signal, due to the lack of frame-wise or sample-wise annotation.
Yet with the stroke-rehab-driven features and trial-wise annotation, simple neural networks such as multi-layer perceptron(MLP) can be applied, and here we \xic{use} a 3-layer MLP.

\begin{table}[ht]
	\newcommand{\tabincell}[2]{\begin{tabular}{@{}#1@{}}#2\end{tabular}}
	\centering
	\begin{tabular}{|c|c|c|}
		\hline
		Predictive Models & \tabincell{c}{RMSE (Acute)}   & {RMSE (Chronic)} \\
		\hline
		Neural Network  & 10.50 & 4.93 \\
		\hline
	    Support vector regression (linear) & 7.47  & 3.25 \\
		\hline
		Support vector regression (rbf) & 9.67  & 4.92 \\
		\hline
		Random forest regression & 8.19    & 3.93  \\
		\hline
		Linear fixed-effects model & 7.24  & 3.29 \\
		\hline
		LMGP  & 5.75   & 3.12 \\
		\hline
	\end{tabular}
	\caption{Predictive Model Comparison based on the proposed stroke-rehab-driven features (in LOSO-CV setting)}
	\label{table:modelComparison}
\end{table}

LOSO-CV \xic{is applied with} the mean RMSE values reported in Table \ref{table:modelComparison}, from which
we \xic{observe} linear models (linear SVR and linear fixed-effects model) yield better results  than non-linear methods (NN, SVR with rbf, and RF).
One of the explanation is the over-fitting effect, where the trained non-linear models do not generalise well to the unseen patients/environments in this longitudinal study setting.
RF is normally known as a classifier with high generalisation capability, yet it may suffer from the low-dimensionality of the selected features (6 features for acute patients and 10 features for chronic patients).
Given the simplicity of the linear models and the designed low-dimensional features, linear models tend to suffer less from the over-fitting effect, with reasonable results in these challenging environments.
\xic{Compared} with linear models, our LMGP can further model the longitudinal mixed-effects (i.e., with linear fixed-effect part and non-linear random-effects part), making the system adaptive to different subjects/time-slots, with the lowest errors.

\begin{table}[ht]
	\newcommand{\tabincell}[2]{\begin{tabular}{@{}#1@{}}#2\end{tabular}}
	\centering
	\begin{tabular}{|c|c|c|}
		\hline
		Methods & \tabincell{c}{RMSE (Acute)}   & {RMSE (Chronic)} \\
		\hline
		Tang et al. \cite{Tang_stroke} & 15.98  & 12.76 \\
        \hline
		Halloran et al. \cite{Shane_ISWC}& 10.12  & 12.14 \\
		\hline
		Ours  & 5.75   & 3.12 \\
		\hline
	\end{tabular}
	\caption{Method comparison (in LOSO-CV setting)}
	\label{table:SOTA}
\end{table}

We also \xic{compare} our approach with other automated CAHAI score regression methods \cite{Tang_stroke} \cite{Shane_ISWC} in the existing literature.
Different from our approach, \cite{Tang_stroke} and \cite{Shane_ISWC} are pure data-driven approaches.
To address the lack of annotation problem, Tang et al. use GMM clustering (on the sliding windows) \cite{Tang_stroke} to learn latent features that can be aggregated for trial-wise representation, 
while Halloran et al. \cite{Shane_ISWC} employ pseudo labelling strategy for trial-wise representation.
However, both data-driven features cannot suppress the substantial noises in the original accelerator signal, and such noises (e.g., irrelevant daily activities) significantly affect the performance of both approaches.
In contrast, by taking advantage of the domain knowledge, our proposed stroke-rehab-driven representation is compact yet informative, and from Table \ref{table:SOTA} and Table \ref{table:modelComparison} we can see it tends to have lower errors than \cite{Tang_stroke} \cite{Shane_ISWC} irrespective of the predictive models for both patient groups.

. 
\section{Conclusions}\label{section:conclusions}
In this work, we \xic{develop} an automated stroke rehabilitation assessment system using wearable sensing and machine learning techniques.
We \xic{collect} accelerometer data using wrist-worn sensors, based on which we \xic{build} models for CAHAI score prediction, which can provide objective and continuous rehabilitation assessment.
To map the long time-series (i.e., 3-day accelerometer data) to the CAHAI score, we \xic{propose} a pipeline which \xic{can perform from} data cleaning, feature design, to predictive model development.
Specifically, we \xic{propose} two compact features which can well capture the rehabilitation characteristics while suppressing the irrelevant daily activities, which is crucial on analysing the data collected in free-living environments.
We further \xic{use} LMGP, which can make the model adaptive to different subjects and different time slots (across different weeks).
Comprehensive experiments \xic{are} conducted on both acute/chronic patients, and very promising results \xic{are} achieved, especially on the chronic patient group.
We also \xic{study} different feature subsets on modelling the fixed-effects part in LMGP, and experiments \xic{suggest} the errors can be further reduced for the challenging acute patient population.

Due to irrelevant daily activities and strong heterogeneity among subjects, it is very challenging for researchers in mathematics, computing sciences and other areas to deal with free-living data. It is also crucial to develop models which have good mathematical properties and have physical explanation particularly in medical research. Hopefully, the ideas of the new features and the models discussed in this paper can provide some hints on addressing similar problems in health research.
\section*{Appendix}\label{section3:app}

\xic{
\subsection{List of Abbreviations/notations}
\begin{itemize}
    \item \textbf{VM}			Signal vector magnitude	
    \item \textbf{DWT}          Discrete wavelet transform
    \item \textbf{DWPT}         Discrete wavelet packet transform
    \item \textbf{LMGP}         Longitudinal mixed-effects Gaussian process prior
    \item \textbf{SAD}          Normalised Sum of Absolute value of the wavelet coefficients at different Decomposition scales
    \item \textbf{PNP}          wavelet features that combine both Paralysed side and Non-Paralysed side
\end{itemize}
}

\subsection{The CAHAI score form}\label{section:CAHAI_form}
\begin{figure}[H]
	\centering
	\includegraphics[width=11cm,height=10cm]{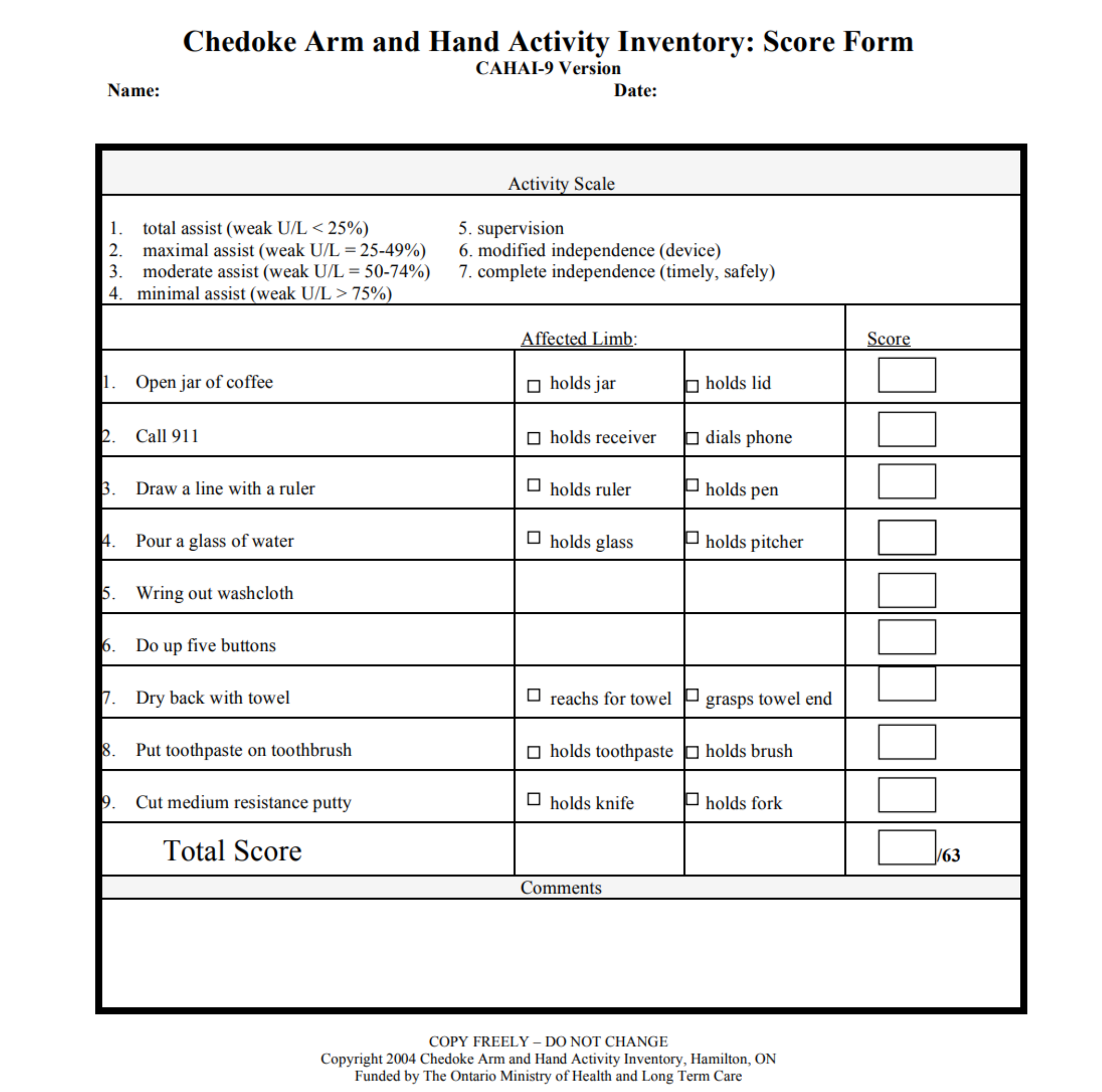}
	\caption{The CAHAI score form \cite{Barreca:2006b}.}
	\label{figure:CAHAI2}
\end{figure}

\subsection{Discrete wavelet transform and discrete wavelet packet transform}\label{section:DWTDWPT}
The \textbf{DWT} procedure includes two parts: decomposition and reconstruction. Decomposition part will be the main focus in this project. We now consider more details of the \textbf{DWT} using matrix algebra:
\begin{equation}\label{equation:DWTraw}
\mathbf{W}=\mathcal{W} \mathbf{X},
\end{equation}
where $\ve{W}$ is the output of matrix of \textbf{DWT} coefficients in different scales. $\ve{W}$ is the orthonormal matrix containing different orthonormal wavelet bases (more details can be checked in \cite{Daubechies:2006} and \cite{waveletbook}) and it satisfies $\mathcal{W}^T\mathcal{W} = \mathbf{I}_N$. $\mathbf{X}$ is the raw signal. The signal $\mathbf{X}$ with length $N = 2^J$, the $N \times N$ orthonormal matrix $\ve{W}$ can be separated into J+1 submatrices, each of which can produce a partitioning of the vector $\mathbf{W}$ of \textbf{DWT} coefficients in each scale j, j = 1,2,..., J.
To be more specific, Eq\eqref{equation:DWTraw} can be rewritten as follows:

\begin{equation}\label{aaa}
\mathcal{W} \mathbf{X} = \begin{bmatrix} \mathcal{W}_1 \\ \mathcal{W}_2 \\ \vdots \\ \mathcal{W}_J \\ \mathcal{V}_J \\ \end{bmatrix} \textbf{X} = \begin{bmatrix} \mathcal{W}_1 \mathbf{X} \\ \mathcal{W}_2 \mathbf{X}\\ \vdots \\ \mathcal{W}_J \mathbf{X}\\ \mathcal{V}_J \mathbf{X}\\ \end{bmatrix} = \begin{bmatrix} \mathbf{W}_1 \\ \mathbf{W}_2 \\ \vdots \\ \mathbf{W}_J \\ \mathbf{V}_J \\ \end{bmatrix} = \mathbf{W},
\end{equation}
where $\mathbf{W}_j$ is a column vector of length $N/{2^j}$ representing the differences in adjacent weighted averages from scale 1 to scale J, $\mathbf{V}_J$ is the last column contained in $\mathbf{W}$ which has the same length with $\mathbf{W}_J$. $\mathbf{W}_j$ is defined as detailed coefficients at scale j. $\mathbf{V}_J$ contains the approximated coefficients at the J-th level. $\mathcal{W}_j$ has dimension $N/{2^j} \times N$, where j = 1,2,...,J and $\mathbf{V}_J$ has the same dimension with $\mathbf{W}_J$. Note that the rows of design orthonormal matrix $\mathcal{W}$ depend on the decomposition level j-th. In other words, the value of J depends on the \textbf{DWT} decomposition scale of the raw signal. The maximum decomposition level j equals J since our signal $\mathbf{X}$ has length $N = 2^J$.

We now further consider wavelet packet transform \textbf{DWPT}. The \textbf{DWPT} is the expansion of the discrete wavelet transformation. In \textbf{DWT}, each scale is calculated by passing only the previous wavelet approximated coefficients through discrete-time low and high pass quadrature mirror filters. However, in the \textbf{DWPT}, both the detailed and approximation coefficients are decomposed to create the full binary tree.  More details can be found in \cite{waveletbook}.


\subsection{Commonly used wavelet features}\label{section:waveletfeatures}

In the discrete wavelet transform (\textbf{DWT}), $ \mathbf{W}_j$ represents \textbf{DWT} coefficients in the j-th decomposition scale. \textbf{DWT} can be written as $\ \mathbf{W}=\mathcal{W}  \mathbf{X}$, where $ \mathbf{W}$ is a column vector with length $2^j$ and $ \mathbf{W} = [ \mathbf{W}_1,  \mathbf{W}_2, ... ,  \mathbf{W}_J, \mathbf{V}_J]^{\mathrm{T}}$, $\mathcal{W}$ is the orthonormal matrix which satisfies $\mathcal{W}^T \mathcal{W} =  \mathbf{I}_n$ and contains different filters. Due to the orthonormality of \textbf{DWT}, which means that $\mathbf{X}=\mathcal{W}^{\mathrm{T}}  \mathbf{W}$ and $\left\| \mathbf{X}\right\|{^2} = \left\|  \mathbf{W} \right\|{^2}$, $\left\| \mathbf{W}_j \right\|{^2}$ shows energy in the \textbf{DWT} coefficients with decomposition level $j$.
Now the energy preserving condition can be written as:
\begin{equation}
\left\| \mathbf{X} \right\|{^2} = \left\| \mathbf{W} \right\|{^2} = \sum_{j=1}^{J} \left\| \mathbf{W}_j \right\|{^2} + \left\|  \mathbf{V}_J \right\|{^2},
\end{equation}
where $\mathbf{X}$ is our VM data (the signal vector magnitude of accelerometer data; see Sec.\ref{section:preprocessSVM}) with length N, $j = 1, 2, ... , J$ is the discrete wavelet transform decomposition level. $\mathbf{W}_j$ denotes the detailed coefficient in scale j, and is a vector of length $N/{2^j}$ representing the differences in adjacent weighted averages from scale 1 to scale J. $\mathbf{V}_J$ denotes the approximated coefficients in the Jth level and has the same length as $\mathbf{W}_J$. Based on the decomposition, each $\left\| \mathbf{W}_j \right\|{^2}$ represents a special part of the energy in our VM data which relates to the certain frequency domain \cite{Preece:2009} \cite{waveletbook}.
Then the sample variance from \cite{waveletbook} can be decomposed as:
\begin{equation}
\widehat{\sigma}_{\mathbf{X}}^2 = \frac{1}{N} \left\| \mathbf{W} \right\|{^2} - \overline{X} =   \sum_{j=1}^{J} \frac{\left\| \mathbf{W}_j \right\|{^2}}{N}.
\label{equation:samplevariance}
\end{equation}
The term $\frac{\left\| \textbf{W}_j \right\|{^2}}{N} $ represents the sample variance (corresponding to $j$ at different scales of \textbf{DWT} decomposition) in our VM data $\mathbf{X}$.

There are many wavelet features (e.g., \cite{Preece:2009}) for the classification of dynamic activities from accelerometer data using \textbf{DWT}. On this basis, we extract the features from the energy preserving condition and sample variance mentioned previously.

{ We aim to look for the features which imply the recovery level among the stroke patients (see Sec.\ref{section:feat_design}).}
Now, we define the features in the j-th level discrete wavelet transform and discrete wavelet packet transform:
\begin{equation*}
	\mathbf{SSD}_{j} = \frac{\left \| \textbf{W}_{j} \right \|^2}{N/{2^j}} = 2^j \frac{\left \| \textbf{W}_{j} \right \|^2}{N}.
\end{equation*}
{ For the detailed coefficients $\textbf{W}_{j}$ at decomposition level j, $\left \| \textbf{W}_{j} \right \|^2$ presents its energy and the raw data with length N}. Hence the physical explanation of  $\mathbf{SSD}_{j}$ is that it stands for the point energy at the decomposition level j.  Moreover, { from the Eq\eqref{equation:samplevariance}, $\frac{ \left\| \textbf{W}_j \right\|{^2}}{N}$ represents the sample variance at the decomposition level j}, $\mathbf{SSD}_{j}$ also has properties of both the energy preserving condition and the sample variance in wavelet analysis with constant ${2^j}$ .

Comparing with $\mathbf{SSD}_{j}$ (sum of Square value of DWT coefficients at scale $j$ (with normalisation)), we define other features call $\mathbf{SAD}_{j}$, which is sum of Absolute value of DWT coefficients at scale $j$ (with normalisation):

\begin{equation*}
	\mathbf{SAD}_{j} = \frac{\left \| \mathbf{W}_{j} \right \|_1}{N/{2^j}} = 2^j \frac{\left \| \mathbf{W}_{j} \right \|_1}{N}.
\end{equation*}

{ After we check the correlation between  the  important wavelet feature $\mathbf{PNP}$ ( Sec.\ref{section:feat_design}) and CAHAI score, the branch of features $\mathbf{PNP}$ using $\mathbf{SAD}$ based perform better than those using $\mathbf{SSD}$ based in Table \ref{table:corrCAHAISADSSD}. Hence we consider the commonly used feature  $\mathbf{SAD}_{j}$ in this paper. }

\begin{table}[ht]
		\newcommand{\tabincell}[2]{\begin{tabular}{@{}#1@{}}#2\end{tabular}}
	\centering
	\begin{tabular}{|c|cccc|cccc|}
		\hline
		-	&  &Acute  &Patients & 	&  &Chronic &Patients & \\
		\hline
		Scale (k)&\tabincell{c}{$PNP^1_k$\\$(SSD)$}  &  \tabincell{c}{$PNP^2_k$\\$(SSD)$} & \tabincell{c}{$PNP^1_k$\\$(SAD)$} & \tabincell{c}{$PNP^2_k$\\$(SAD)$} & \tabincell{c}{$PNP^1_k$\\$(SSD)$} &  \tabincell{c}{$PNP^2_k$\\$(SSD)$} & \tabincell{c}{$PNP^1_k$\\$(SAD)$} & \tabincell{c}{$PNP^1_k$\\$(SAD)$} \\
		\hline
		k=1.1 & 0.60 & -0.65 & 0.68 & -0.70  & 0.45 & -0.45 & 0.56 & -0.56  \\
		k=1.2 & 0.60 & -0.66 & 0.69 & -0.71  & 0.46 & -0.45 & 0.57 & -0.56 \\
		k=1.3 & 0.63 & -0.69 & 0.70 & -0.72 & 0.49 & -0.48 & 0.58 & -0.57 \\
		k=1.4 & 0.62 & -0.68 & 0.69 & -0.71  & 0.47 & -0.47 & 0.57  & -0.57 \\
		k=2   & 0.65 & -0.69 & 0.69 & -0.71  & 0.45 & -0.45 & 0.56  & -0.55 \\
		k=3   & 0.63 & -0.67 & 0.67 & -0.68  & 0.39 & -0.38 & 0.53  & -0.52 \\
		k=4   & 0.59 & -0.63 & 0.60 & -0.63  & 0.31 & -0.30 & 0.48  & -0.47 \\
		k=5   & 0.46 & -0.50 & 0.49 & -0.52  & 0.29 & -0.27 & 0.43  & -0.42 \\
		k=6   & 0.32 & -0.38 & 0.35 & -0.38  & 0.20 & -0.16 & 0.35  & -0.34 \\
		k=7   & 0.16 & -0.19 & 0.19 & -0.20  & 0.13 & -0.10 & 0.25  & -0.24 \\
		\hline
	\end{tabular}
	\caption{The correlation between SAD and SSD based wavelet features and CAHAI score for acute and chronic patients .}
	\label{table:corrCAHAISADSSD}
\end{table}

In our analysis, we assume the discrete wavelet decomposition level $J=7$ which is the same level as in \cite{Sekine:1998} and contains enough low-frequency component as the stroke patients' movement. The frequency domain with seven scales is shown in Table \ref{Table:scale-frequency}:
\begin{table}[ht]
	\centering
	\begin{tabular}{|c|c|c|c|}
		\hline
		& Scale 7  & Scale 6 & Scale 5  \\
		\hline
		Frequency & 0.0078hz-0.0156hz & 0.0156hz - 0.0312hz & 0.0312hz - 0.0625hz   \\
		\hline
		& Scale 4  & Scale 3  & Scale 2   \\
		\hline
		Frequency & 0.0625hz - 0.125hz & 0.125hz - 0.25hz&  0.25hz - 0.50h  \\
		\hline
		& Scale 1  &   &    \\
		\hline
		Frequency & 0.50hz - 1hz & &  \\
		\hline		
	\end{tabular}
	\caption{The frequency domain from scale 1 to scale 7 by using DWT. }
	\label{Table:scale-frequency}
\end{table}

So far, we have decomposed the VM data $\textbf{X}$ to get $\textbf{W}_1$, $\textbf{W}_2$, ... , $\textbf{W}_7$ using \textbf{DWT}. { Since the frequency domain at scale 1 is so wide (0.50hz - 1hz), it is better to divide it into smaller one},  then using \textbf{DWPT} in Appendix \ref{section:DWTDWPT}, we can further decompose  $\textbf{W}_1$ into $\textbf{W}_{3.4}$, $\textbf{W}_{3.5}$, $\textbf{W}_{3.6}$ and $\textbf{W}_{3.7}$ which are the results of the $3$-rd stage of \textbf{DWPT}, each coefficient vector with length $N/{2^3}$ has the same dimension as the coefficients in the third level of \textbf{DWT} decomposition, that is
$$
\left\| \textbf{X} \right\|{^2} = \left\| \textbf{W} \right\|{^2} = \left\| \textbf{W}_{3.4} \right\|{^2} + \left\| \textbf{W}_{3.5} \right\|{^2} + \left\| \textbf{W}_{3.6} \right\|{^2} + \left\| \textbf{W}_{3.7} \right\|{^2}  + \sum_{j=2}^{J} \left\| \textbf{W}_j \right\|{^2} + \left\|  \textbf{V}_J \right\|{^2}.
$$

Now we have coefficients at 10 decomposition scales by using \textbf{DWT} and \textbf{DWPT}: $ \textbf{W}_{3.4}$, $\textbf{W}_{3.5}$, $\textbf{W}_{3.6}$,  $\textbf{W}_{3.7}$, $\textbf{W}_2$, $\textbf{W}_3$, $\textbf{W}_4$, $\textbf{W}_5$, $\textbf{W}_6$ and $\textbf{W}_7$. Based on these detailed coefficients, we define the commonly used wavelet features again:

\begin{equation*}
	\begin{split}
		\textbf{Scale 1.1}: \ & SAD_{1.1} = \frac{\left \| \textbf{W}_{3.4} \right \|_1}{N/{2^3}} = 2^3 \frac{\left \| \textbf{W}_{3.4} \right \|_1}{N}, \\
		\textbf{Scale 1.2}: \ & SAD_{1.2} = \frac{\left \| \textbf{W}_{3.5} \right \|_1}{N/{2^3}} = 2^3 \frac{\left \| \textbf{W}_{3.5} \right \|_1}{N}, \\
		\textbf{Scale 1.3}: \ & SAD_{1.3} = \frac{\left \| \textbf{W}_{3.6} \right \|_1}{N/{2^3}} = 2^3 \frac{\left \| \textbf{W}_{3.6} \right \|_1}{N}, \\
		\textbf{Scale 1.4}: \ & SAD_{1.4} = \frac{\left \| \textbf{W}_{3.7} \right \|_1}{N/{2^3}} = 2^3 \frac{\left \| \textbf{W}_{3.7} \right \|_1}{N}, \\
		\textbf{Scale \ j}: \ & SAD_{j} = \frac{\left \| \textbf{W}_{j} \right \|_1}{N/{2^j}} = 2^j \frac{\left \| \textbf{W}_{j} \right \|_1}{N}, \qquad \qquad j = 2,3,4,5,6,7. \\
	\end{split}
\end{equation*}

There are 10 features which provide reliable and valid information { (corresponding to more frequency domains)} from different frequency domains. The frequency domain of these features, among 10 scales, is listed in Table \ref{figure:frequency10}:
\begin{table}[ht]
	\centering
	\begin{tabular}{c|c|c|c}
		\hline
		& Scale 1.1 & Scale 1.2 & Scale 1.3  \\
		\hline
		Frequency & 0.5hz - 0.625hz & 0.625hz - 0.75hz & 0.75hz - 0.875hz  \\
		\hline
		& Scale 1.4 & Scale 2 & Scale 3 \\
		\hline
		Frequency & 0.875hz - 1hz &  0.25-0.50hz & 0.125hz - 0.25hz\\
		\hline
		& Scale 4 & Scale 5& Scale 6 \\
		\hline
		Frequency& 0.0625hz - 0.125hz & 0.0312hz - 0.0625hz  & 0.0156hz - 0.0312hz\\
		\hline
		& Scale 7 & &  \\
		\hline
		Frequency& 0.0078hz - 0.0156hz  &   & \\
		\hline
	\end{tabular}
	\caption{The frequency domain  from scale 1.1 to scale 7 by using DWPT and DWT. }
	\label{figure:frequency10}
\end{table}

\subsection{Performance of LMGP through three different kernels} \label{section:LMGP_kernels} \hfill\\
Three kernels \xic{are} used in LMGP, and they are linear kernel, powered exponential kernel and rational quadratic kernel.
 We \xic{use} the selected features (from Table \ref{table:variableselection}) as the fixed-effects features and random-effects
features, and the results \xic{are} reported in Table \ref{table:kernalscomparison}.
\begin{table}[ht]
		\newcommand{\tabincell}[2]{\begin{tabular}{@{}#1@{}}#2\end{tabular}}
		\centering
		\begin{tabular}{|c|c|c|}
			\hline
			Selected kernels in LMGP& \tabincell{c}{RMSE (Acute)}   & {RMSE (Chronic)} \\
			\hline
			linear kernel & 5.89  & 3.13 \\
			\hline
			powered exponential kernel & 5.75  & 3.12 \\
			\hline
			rational quadratic kernel & 7.58    & 3.24  \\
			\hline
		\end{tabular}
		\caption{Performance of LMGP based on three kernels}
		\label{table:kernalscomparison}
	\end{table}

\bibliographystyle{unsrt}
\bibliography{ref}

\begin{thebibliography}{10}

\bibitem{Donnan:2008}
G.~Donnan, M.~Fisher, M.~Macleod, and S.~Davis.
\newblock {Stroke.}
\newblock {\em Lancet}, 371(2):1612--1623, 2008.

\bibitem{BrainImaging_2005}
M.~Wintermark, M.~Sesay, E.~Barbier, K.~Borbély, W.P. Dillon, J.D. Eastwood,
  T.C. Glenn, C.B. Grandin, S.~Pedraza, J.F. Soustiel, T.~Nariai, G.~Zaharchuk,
  J.M. Caillé, V.~Dousset, and H.~Yonas.
\newblock Comparative overview of brain perfusion imaging techniques.
\newblock {\em Journal of Neuroradiology}, 32(5):294--314, 2005.

\bibitem{Questionaires_2007}
Pietro Ferrari, Christine Friedenreich, and Charles Matthews.
\newblock The role of measurement error in estimating levels of physical
  activity.
\newblock {\em American journal of epidemiology}, 166:832--40, 11 2007.

\bibitem{CAHAI_2005}
Susan~R Barreca, Paul~W. Stratford, Cynthia~L. Lambert, Lisa~M. Masters, and
  David~L Streiner.
\newblock Test-retest reliability, validity, and sensitivity of the chedoke arm
  and hand activity inventory: a new measure of upper-limb function for
  survivors of stroke.
\newblock {\em Archives of physical medicine and rehabilitation}, 86
  8:1616--22, 2005.

\bibitem{Barreca:2005}
S.~Barreca, P.~Stratford, C.~Lambert, L.~Masters, and D.~Streiner.
\newblock {Test-Retest Reliability, Validity, and Sensitivity of the Chedoke
  Arm and Hand Activity Inventory: a New Measure of Upper-Limb Function for
  Survivors of Stroke}.
\newblock {\em Arch Phys Med Rehabil}, 86:1616--1622, 2005.

\bibitem{waveletbook}
Andrew T.~Walden Donald B.~Percival.
\newblock {\em Wavelet methods for time series analysis}.
\newblock Cambridge Series in Statistical and Probabilistic Mathematics.
  Cambridge University Press, 1 edition, 2000.

\bibitem{Preece:2009}
S.~J. Preece*, J.~Y. Goulermas, L.~P.~J. Kenney, and D.~Howard.
\newblock A comparison of feature extraction methods for the classification of
  dynamic activities from accelerometer data.
\newblock {\em IEEE Transactions on Biomedical Engineering}, 56(3):871--879,
  March 2009.

\bibitem{Barreca:2006b}
S.~Barreca, P.~Stratford, L.~Masters, C.~Lambert, J.~Griffiths, and C.~McBay.
\newblock {Validation of Three Shortened Versions of the Chedoke Arm and Hand
  Activity Inventory}.
\newblock {\em Physiother. Can.}, 58:1--9, 2006.

\bibitem{PD_Rehman19}
Rana zia~ur Rehman, Silvia Din, Yu~Guan, Alison Yarnall, Jian Shi, and Lynn
  Rochester.
\newblock Selecting clinically relevant gait characteristics for classification
  of early parkinson's disease: A comprehensive machine learning approach.
\newblock {\em Scientific Reports}, 9, 12 2019.

\bibitem{Nils_PD15}
Nils~Y. Hammerla, James~M. Fisher, Peter Andras, Lynn Rochester, Richard
  Walker, and Thomas Ploetz.
\newblock Pd disease state assessment in naturalistic environments using deep
  learning.
\newblock In {\em Proceedings of the Twenty-Ninth AAAI Conference on Artificial
  Intelligence}, AAAI, pages 1742--1748. AAAI Press, 2015.

\bibitem{TP_autism}
Thomas Ploetz, Nils~Y. Hammerla, Agata Rozga, Andrea Reavis, Nathan Call, and
  Gregory~D. Abowd.
\newblock Automatic assessment of problem behavior in individuals with
  developmental disabilities.
\newblock In {\em Proceedings of the 2012 ACM Conference on Ubiquitous
  Computing}, UbiComp, pages 391--400, New York, NY, USA, 2012. Association for
  Computing Machinery.

\bibitem{Little_depression2020}
Bethany Little, Ossama Alshabrawy, Daniel Stow, I.~Ferrier, Roisin McNaney,
  Daniel Jackson, Karim Ladha, Cassim Ladha, Thomas Ploetz, Jaume Bacardit,
  Patrick Olivier, Peter Gallagher, and John O'Brien.
\newblock Deep learning-based automated speech detection as a marker of social
  functioning in late-life depression.
\newblock {\em Psychological Medicine}, pages 1--10, 01 2020.

\bibitem{Bing_sleep}
Bing Zhai, Ignacio Perez-Pozuelo, Emma A.~D. Clifton, Joao Palotti, and
  Yu~Guan.
\newblock Making sense of sleep: Multimodal sleep stage classification in a
  large, diverse population using movement and cardiac sensing.
\newblock {\em Proc. ACM Interact. Mob. Wearable Ubiquitous Technol.}, 4(2),
  June 2020.

\bibitem{Yike_sleep}
Akara Supratak, Hao Dong, Chao Wu, and Yike Guo.
\newblock Deepsleepnet: a model for automatic sleep stage scoring based on raw
  single-channel eeg.
\newblock {\em IEEE Transactions on Neural Systems and Rehabilitation
  Engineering}, PP, 03 2017.

\bibitem{Yang_ISWC}
Yang Bai, Yu~Guan, and Wan-Fai Ng.
\newblock Fatigue assessment using ecg and actigraphy sensors.
\newblock In {\em Proceedings of the 24rd International Symposium on Wearable
  Computers}, ISWC, New York, NY, USA, 2020. Association for Computing
  Machinery.

\bibitem{Fatigue_IMU}
Alzhraa~A Ibrahim, Arne Küderle, Heiko Gaßner, Jochen Klucken, Bjoern~M
  Eskofier, and Felix Kluge.
\newblock Inertial sensor-based gait parameters reflect patient-reported
  fatigue in multiple sclerosis.
\newblock {\em Journal of neuroengineering and rehabilitation}, 17(1):165,
  December 2020.

\bibitem{Anna_IMU}
AM~Ratcliffe, B~Zhai, Y~Guan, D~Jackson, SWARM, and JR. Sneyd.
\newblock Patient-centred measurement of recovery from day-case surgery using
  wrist worn accelerometers: a pilot and feasibility study.
\newblock {\em Anaesthesia}, 2020.

\bibitem{gait_surgery}
Reed Gurchiek, Rebecca Choquette, Bruce Beynnon, James Slauterbeck, Timothy
  Tourville, Michael Toth, and Ryan McGinnis.
\newblock Open-source remote gait analysis: A post-surgery patient monitoring
  application.
\newblock {\em Scientific reports}, 9:17966, 11 2019.

\bibitem{Dolatabadi:2017}
Elham Dolatabadi, Ying Zhi, Bing Ye, Marge Coahran, Giorgia Lupinacci, Alex
  Mihailidis, Rosalie Wang, and Babak Taati.
\newblock The toronto rehab stroke pose dataset to detect compensation during
  stroke rehabilitation therapy.
\newblock pages 375--381, 05 2017.

\bibitem{sensing_sEMG}
A.~C. {Ganesh}, B.~S. {Renganathan}, C.~{Rajakumaran}, S.~P. {Preejith},
  K.~{Shubham}, J.~{Jayaraj}, and S.~{Mohanasankar}.
\newblock Post-stroke rehabilitation monitoring using wireless surface
  electromyography: A case study.
\newblock In {\em 2018 IEEE International Symposium on Medical Measurements and
  Applications (MeMeA)}, pages 1--6, 2018.

\bibitem{sensing_5sensors}
H.~{Jung}, J.~{Park}, J.~{Jeong}, T.~{Ryu}, Y.~{Kim}, and S.~I. {Lee}.
\newblock A wearable monitoring system for at-home stroke rehabilitation
  exercises: A preliminary study.
\newblock In {\em 2018 IEEE EMBS International Conference on Biomedical Health
  Informatics (BHI)}, pages 13--16, 2018.

\bibitem{sensing_ecosystem}
Maxence Bobin, Franck Bimbard, Mehdi Boukallel, Margarita Anastassova, and
  Mehdi Ammi.
\newblock Spectrum: Smart ecosystem for stroke patient's upper limbs
  monitoring.
\newblock 13, 02 2019.

\bibitem{Shane_ISWC}
Shane Halloran, Lin Tang, Yu~Guan, Jian~Qing Shi, and Janet Eyre.
\newblock Remote monitoring of stroke patients' rehabilitation using wearable
  accelerometers.
\newblock In {\em Proceedings of the 23rd International Symposium on Wearable
  Computers}, ISWC 19, pages 72--77, New York, NY, USA, 2019. Association for
  Computing Machinery.

\bibitem{Tang_stroke}
Lin Tang, Shane Halloran, Jian~Qing Shi, Yu~Guan, Chunzheng Cao, and Janet
  Eyre.
\newblock Evaluating upper limb function after stroke using the free-living
  accelerometer data.
\newblock {\em Statistical Methods in Medical Research}, 2020.

\bibitem{Shi:2012}
J.Q. Shi, B.~Wang, E.J. Will, and R.M. West.
\newblock Mixed-effects {Gaussian} process functional regression models with
  application to dose response curve prediction.
\newblock {\em Statistics in Medicine}, 31(26):3165--3177, 2012.

\bibitem{Shi:2013}
J.Q. Shi, Y.~Cheng, J.~Serradilla, G.~Morgan, C.~Lambden, G.~Ford, C.~Price,
  H.~Rodgers, T.~Cassidy, L.~Rochester, and J.A. Eyre.
\newblock {Evaluating Functional Ability of Upper Limbs after Stroke Using
  Video Game Data}.
\newblock In K.~Imamura, S.~Usui, T.~Shirao, T.~Kasamatsu, L.~Schwabe, and
  N.~Zhong, editors, {\em International Conference on Brain and Health
  Informatics}, volume 8211 of {\em Lecture Notes in Artificial Intelligence},
  pages 181--192. Springer, 2013.

\bibitem{Axivity}
{Axivity Ltd}.
\newblock {AX3, 3-Axis Logging Accelerometer}.
\newblock \url{https://axivity.com/product/ax3}.
\newblock [Online; accessed July-2020].

\bibitem{ax3_UKB}
Aiden Doherty, Dan Jackson, Nils Hammerla, Thomas Ploetz, Patrick Olivier,
  Malcolm~H. Granat, Tom White, Vincent~T. van Hees, Michael~I. Trenell,
  Christoper~G. Owen, Stephen~J. Preece, Rob Gillions, Simon Sheard, Tim
  Peakman, Soren Brage, and Nicholas~J. Wareham.
\newblock Large scale population assessment of physical activity using wrist
  worn accelerometers: The uk biobank study.
\newblock {\em PLOS ONE}, 12(2):1--14, 02 2017.

\bibitem{Bouten:1997}
Carlijn Bouten, Karel Koekkoek, Maarten Verduin, Rens Kodde, and Jan Janssen.
\newblock A triaxial accelerometer and portable data processing unit for the
  assessment of daily physical activity.
\newblock {\em IEEE transactions on bio-medical engineering}, 44:136--47, 04
  1997.

\bibitem{Guan_HAR17}
Yu~Guan and Thomas Ploetz.
\newblock Ensembles of deep lstm learners for activity recognition using
  wearables.
\newblock {\em Proc. ACM Interact. Mob. Wearable Ubiquitous Technol.}, 1(2),
  June 2017.

\bibitem{HAR_TP}
T.~Ploetz and Y.~Guan.
\newblock Deep learning for human activity recognition in mobile computing.
\newblock {\em Computer}, 51(5):50--59, 2018.

\bibitem{Henrik:1999}
Henrik~[Stig JÃ¸rgensen], Hirofumi Nakayama, Hans~Otto Raaschou, and
  Tom~[SkyhÃ¸j Olsen].
\newblock Stroke: Neurologic and functional recovery the copenhagen stroke
  study.
\newblock {\em Physical Medicine and Rehabilitation Clinics of North America},
  10(4):887 -- 906, 1999.
\newblock A New Century Approach to Stroke Management and Rehabilitation.

\bibitem{acc_SVM}
D.~M. {Karantonis}, M.~R. {Narayanan}, M.~{Mathie}, N.~H. {Lovell}, and B.~G.
  {Celler}.
\newblock Implementation of a real-time human movement classifier using a
  triaxial accelerometer for ambulatory monitoring.
\newblock {\em IEEE Transactions on Information Technology in Biomedicine},
  10(1):156--167, 2006.

\bibitem{babystroke}
Yan Gao, Yang Long, Yu~Guan, Anna Basu, Jessica Baggaley, and Thomas Ploetz.
\newblock Towards reliable, automated general movement assessment for perinatal
  stroke screening in infants using wearable accelerometers.
\newblock {\em Proc. ACM Interact. Mob. Wearable Ubiquitous Technol.}, 3(1),
  March 2019.

\bibitem{wavelet_IMU}
Fouaz~S Ayachi, Hung~P Nguyen, Catherine Lavigne-Pelletier, Etienne Goubault,
  Patrick Boissy, and Christian Duval.
\newblock Wavelet-based algorithm for auto-detection of daily living activities
  of older adults captured by multiple inertial measurement units (imus).
\newblock {\em Physiological measurement}, 37(3):442--461, March 2016.

\bibitem{shi:2011}
Jian Shi and Taeryon Choi.
\newblock {\em Gaussian Process Regression Analysis for Functional Data}.
\newblock London: Chapman and Hall/CRC, 01 2011.

\bibitem{Daubechies:2006}
I.~Daubechies.
\newblock Orthonormal bases of compactly supported wavelets.
\newblock {\em Commun. Pure. Appl. Math}, pages 909--996, 2006.

\bibitem{Sekine:1998}
M.~Sekine, T.~Tamura, M.~Ogawa, T.~Togawa, and Y.~Fukui.
\newblock Classification of acceleration waveform in a continuous walking
  record.
\newblock In {\em Proceedings of the 20th Annual International Conference of
  the IEEE Engineering in Medicine and Biology Society. Vol.20 Biomedical
  Engineering Towards the Year 2000 and Beyond (Cat. No.98CH36286)}, volume~3,
  pages 1523--1526 vol.3, Oct 1998.

\end{thebibliography}


\end{document}